\documentclass[useAMS,usenatbib]{mn2e}

\usepackage{amsmath}
\usepackage{amssymb}   
\usepackage{amscd}
\usepackage[english]{babel}
\usepackage{graphicx}
\usepackage{bm}
\usepackage{textcomp}
\usepackage{mathrsfs}
\usepackage{tikz}
\usepackage{booktabs}
\newcommand{\x}{\mathbf{x}}

\renewcommand{\hat}{\widehat}

\newcommand{\dertot}[2]{\frac{d #1}{d #2}}

\def\ros#1{{\scriptsize \textrm{#1}}}
\title[Juno Radio Science Experiment]{On the Juno Radio Science Experiment:
  models, algorithms and sensitivity analysis}

\author[G. Tommei et al.]{G. Tommei$^1$\thanks{E-mail:
    tommei@dm.unipi.it}, L. Dimare$^2$\thanks{E-mail:
    dimare@spacedys.com}, D. Serra$^1$\thanks{E-mail:
    dserra@mail.dm.unipi.it}, and A. Milani$^1$\thanks{E-mail: milani@dm.unipi.it}\\
    $^1$ Department of Mathematics, University of Pisa, Pisa, Italy \\
    $^2$ SpaceDyS srl, Cascina (Pisa), Italy}

\begin{document}
\date{Accepted . Received ; in original form }

\maketitle
\label{firstpage}

\begin{abstract}
Juno is a NASA mission launched in 2011 with the goal of studying
Jupiter. The probe will arrive to the planet in 2016 and will be placed
for one year in a polar high-eccentric orbit to study the composition
of the planet, the gravity and the magnetic field.  The Italian Space
Agency (ASI) provided the radio science instrument KaT (Ka-Band
Translator) used for the gravity experiment, which has the goal of
studying the Jupiter's deep structure by mapping the planet's gravity:
such instrument takes advantage of synergies with a similar tool in
development for BepiColombo, the ESA cornerstone mission to
Mercury. The Celestial Mechanics Group of the University of Pisa,
being part of the Juno Italian team, is developing an orbit
determination and parameters estimation software for processing the
real data independently from NASA software ODP. This paper has a
twofold goal: first, to tell about the development of this software
highlighting the models used, second, to perform a sensitivity
analysis on the parameters of interest to the mission.
\end{abstract}


\section{Introduction}\label{sec:intro}

The Juno mission is part of the NASA New Frontiers Program and its
primary goal is the understanding of the origin and evolution of the
planet Jupiter. It has been launched from Cape Canaveral on August 5,
2011 and it will complete its journey in about five years arriving to
Jupiter in 2016, after having experienced an Earth fly-by in 2013.

\begin{figure}
\begin{center}
\includegraphics[width=6cm]{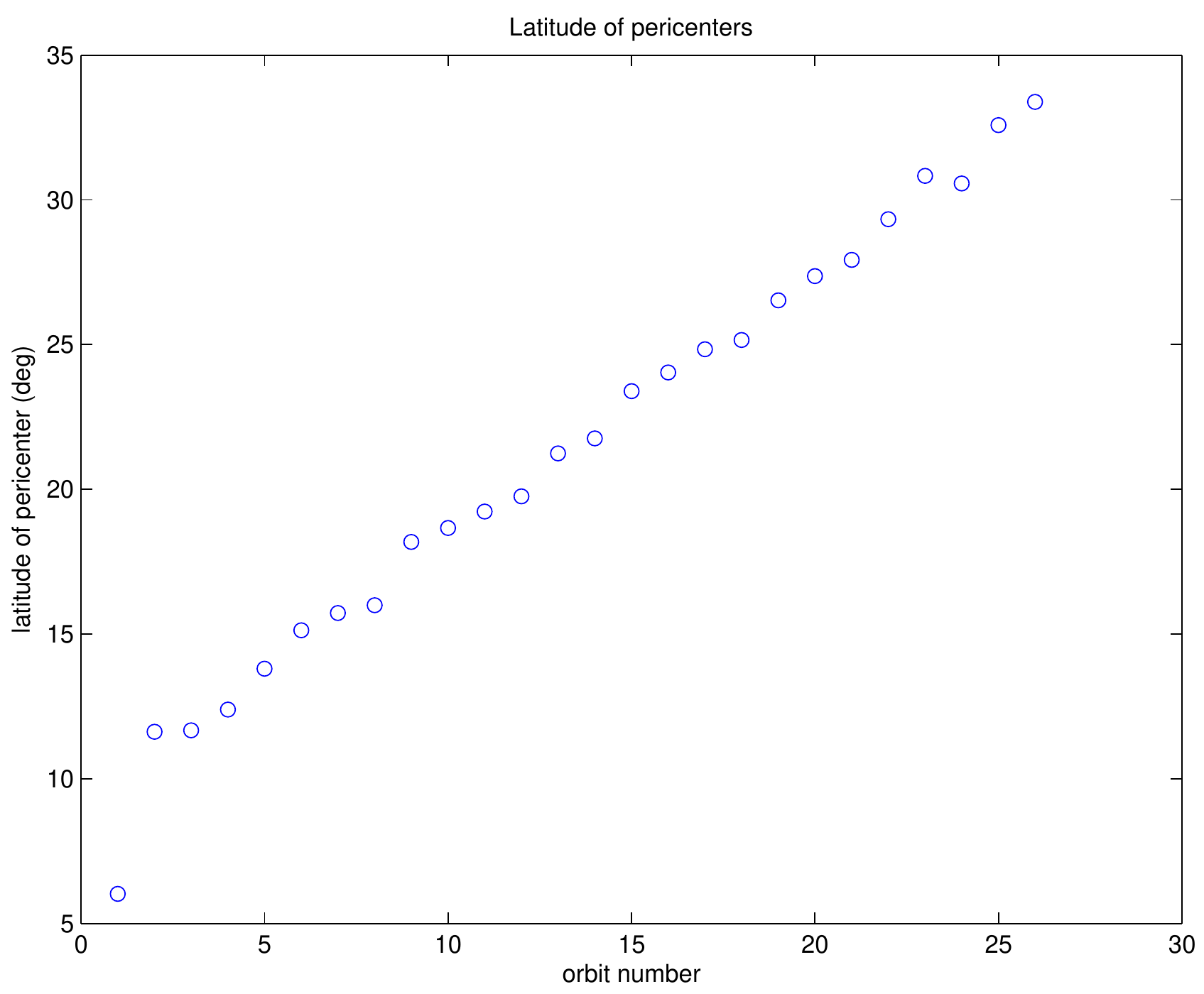}
\end{center}
\label{fig:perijove}
\caption{Perijove latitude variation during the $32$ orbits of Juno.}
\end{figure}

At the end of the five year journey, the spacecraft (S/C) will be inserted in
a high-eccentric polar orbit with a period of about 11 days: the
perijove will be at $\sim 1.06$ Jupiter Radii and apojove at $\sim 39$
Jupiter Radii. The polar orbit with close perijove allows the
spacecraft to avoid the bulk of the Jovian radiation field. During the
nominal mission time, one year, 32 orbits will be explored and $25$ of them will be
dedicated to gravity science. Because of
$J_2$ perturbation by Jupiter the latitude of perijove will change from
$4$ degrees to $34$ degrees (see Figure \ref{fig:perijove}).

Juno's activities will focus on essentially four aspects (see
\cite{MJUNO} for more details):
\begin{itemize}
\item {\bf origin}, measuring the global oxygen abundance, Juno will
  discriminate among different Jupiter's formation scenarios;
\item {\bf interior}, Juno will investigate the structure and
  convection of the Jupiter's interior and through gravity will
  explore the distribution of mass inside the planet;
\item {\bf atmosphere}, the mission will provide three-dimensional
  views of the atmosphere in order to answer to basic questions about
  circulation;
\item {\bf magnetosphere}, Juno will measure signatures of different
  auroral processes in order to study the magnetic field.
\end{itemize}

The science goals of the Juno mission will be achieved with nine
instruments on board the spacecraft. Among these, there is the
communication subsystem to perform Doppler tracking, formed by the
X-Band and Ka-Band transponders interfacing with the 34-meter Deep
Space Network antenna at Goldstone, California. The Italian Space
Agency (ASI) contributed to the Juno mission providing the Ka-Band
Translator (KaT) (developed by the University of Rome ``La Sapienza''
and Thales Alenia Space) used for the gravity experiment. Such
instrument takes advantage of synergies with a similar tool in
development for BepiColombo, the ESA cornerstone mission to Mercury
(see \cite{IB} for more details).

The Celestial Mechanics Group of the University of Pisa is responsible
for the development of an Orbit Determination (OD) and Parameters
Estimation (PE) software for processing the real data independently
from the NASA ones (former ODP and now MONTE); in particular, a gravimetry and a rotation
experiment will be carried out by means of very accurate range-rate
(doppler) observations (the accuracy should be $3 \times 10^{-4}$ cm/s
in range-rate, at 1000 s integration time).

Every time we have to develop a complicated software to process accurate
tracking data, we have to do a preliminary
work consisting in projecting the architecture of the program and in choosing
the appropriate dynamical models. The aim of this paper is to describe this
preliminary, but necessary, work and the description is divided in two parts 
essentially: in the first, we are going to describe the architecture of the software 
for the Juno Radio Science Experiment (RSE) and its state of the art, highlighting the
theoretical models used and the problems addressed; in the second, we
are going to perform a sensitivity analysis on the parameters of
interest to the mission, in order to hypothesize what could be the
results, in terms of parameters accuracies, of the RSE. These analysis is the basis
of the future development of a differential corrector that should estimate all
the interested scientific parameters.

The paper is organized as follows. In Sec.~\ref{sec:gs}, we introduce
fundamental definitions explaining the basics of the OD methodology
used and we describe the global structure of the software pointing
out the main architectural choices done in the development phase. In
Sec.~\ref{sec:dyn} we discuss the dynamics involved in the computation
of the observables, while in Sec.~\ref{sec:obs} we underline the
light-time process and the numerical methods used to avoid
rounding-off errors. Sec.~\ref{sec:sim} is devoted to the sensitivity
analysis of some parameters of interest and to the discussion of the
results of such analysis in terms of scientific goals of the
mission. In Sec.~\ref{sec:future} we draw some conclusions and outline
future developments.

\section{Architectural structure of the OD and PE software for the Juno RSE}\label{sec:gs}

This section describes the architectural choices of the software
system we are developing for the Juno RSE. The main choice done was to
write a \emph{no compromise} software. That is, no approximation, no
shortcut, no choice done just to save time in the development phase,
which may prove a mistake later. All approximations have been
justified, that is tested and certified: this is the only
way to guarantee that the results will be state of the art in all
respects.

As an example, the software needs to include all the parameters which
might affect the observables at the level of accuracy corresponding to
the quality of the measurements (see Sec.~\ref{sec:sim}). In
principle, all of them could be the object of determination in a
global least squares fit to the observables.  Note that the design
choice is to have the possibility of solving for all the unknown
parameters as necessary in one step, as a global least squares fit. It
is also possible to include information from other experiments by
means of {\em a priori observations}, weighted by their normal matrix,
but all the data from the RSE can be directly fit.

However, it is easy to show that some parameters which indeed affect
the data are already known at a level such that they cannot be
improved with the data that will be available. The simplest example is
the orbit of the Moon: it certainly affects the distance between the
ground station and the S/C, since the Earth moves by $\simeq 4,700$ km
as a result of the motion of the Moon. Nevertheless, observing a
satellite around Jupiter is certainly not a good way to constrain the
orbit of the Moon: Lunar Laser Ranging (LLR) already provides the
geocentric orbit of the Moon with an accuracy of few cm, thus the
reflex motion of the Earth around the Earth-Moon barycenter to better
than $1$ mm. A similar argument applies to the satellites of Jupiter:
the satellites, especially the four Galilean ones, have a significant
effect on the orbit of Juno, still the uncertainty in these
perturbations is small. The Juno S/C will in fact neither accurately
measure the positions of the satellites nor pass close enough to them
to be able to measure their gravitational attraction with useful
accuracy. This implies that some of the parameters, including those
contained in the ephemerides of the Moon and other planets, in the
rotation state of the Earth, in the ephemerides of Jupiter's
satellites, and others, do appear in the observables but have to be
handled as \emph{consider parameters}, not to be determined, after
checking that their present uncertainty is such that their
contribution to the uncertainty of the observable is negligible.

\subsection{OD background}\label{sec:backg}

From the algorithmic point of view, the software consists in a non
linear least square differential correction fit of the doppler
tracking data in order to determine the following quantities,
generally referred to as $\mathbf{u}$:
\begin{itemize}
\item[1)] initial conditions of the S/C at given times; 
\item[2)] spherical harmonics of the gravity field of Jupiter according to following
potential ($R$
is the radius of Jupiter)
 \begin{eqnarray*}
 U(r,\theta,\lambda) &= & \frac{GM}{r}\sum_{\ell=0}^{+\infty}\sum_{m=0}^{\ell}\frac{R^{\ell}}
 {r^{\ell}}[C_{\ell m}\cos(m\lambda)+ \\
 && S_{\ell m}\sin(m\lambda)]P_{\ell m}(\sin\theta) \,\, ;
 \end{eqnarray*}
\item[3)] Love's number for the tidal deformation of Jupiter (static tidal theory),
according to the following potential
 \[
 U_{\textrm{Love}}=\sum_{\ell=2}^{+\infty}k_{\ell}\frac{GM_PR^{2\ell+1}}{r_P^{\ell+1}r^{\ell+1}}
 P_{\ell}(\cos\varphi) \,\, ;
 \] 
\item[4)] parameters defining the model of the Jupiter's rotation;
\item[5)] the angular momentum of Jupiter through the Lense-Thirring (LT) effect
causing an acceleration 
\[ \textbf{a}_{LT}=\frac{2\,G}{c^2\,r^3}\,\left [-\textbf{J} \times \dot{\textbf r}+
 3\,\frac{(\textbf{J} \cdot \textbf{r})\,(\textbf{r} \times \dot{\textbf r})}{r^2}  \right ] \,\, ;
 \]
\item[5)] initial conditions for the Barycenter of the Jovian System (BJS) at some initial time; 
\item[6)] relativistic Post-Newtonian Parameter (PPN) $\gamma$ during the
  Superior Conjuction Experiment (SCE).
\end{itemize}

Of course a good initial guess for each of the above parameters will
be also necessary: the navigation team should supply the initial
conditions of the S/C, the gravity and physical parameters will be
taken by previous investigations, the initial conditions for BJS
will be provided by JPL Ephemerides DE421 (see \cite{FWB}) and for
$\gamma$ will be used the General Relativity (GR) value $1$.

Following a classical approach (see, for instance, \cite{MG}), the non
linear least squares fit aims at computing a set of parameters
$\mathbf{u}^*$ which minimizes the target function:
\begin{equation}\label{eq:1}
Q(\mathbf{u})=\frac{1}{m}\boldsymbol{\xi}^T(\mathbf{u})
W\boldsymbol{\xi}(\mathbf{u})=\frac{1}{m}\sum_{i=1}^mw_i\xi_i^2(\mathbf{u}),
\end{equation}
where $m$ is the number of observations and
$\boldsymbol{\xi}=\mathcal{O}-\mathcal{C}$ is the vector of {\it
  residuals}, difference between the observed quantities $\mathcal{O}$
and the predicted ones $\mathcal{C}(\mathbf{u})$, computed following
suitable models and assumptions (described in the next
sections). $\mathcal{O}$ are range and/or range-rate data, while
$\mathcal{C}(\mathbf{u})$ are the results of the light-time
computations (see Sec.~\ref{sec:obs} and \cite{TMV}) as a function of
all the quantities $\mathbf{u}$ listed above. Finally, $w_i$ is the
weight associated to the $i-$observation.

Among the parameters $\mathbf{u}$, 1), 2) and 3) are present in the
equations of motion for the jovicentric orbit of the spacecraft, while
4) and 5) in those for the barycentric orbit of BJS (see
Sec.~\ref{sec:dyn}).

Other information required to such orbit propagations are supposed to
be known: position and velocity for the other planets of the Solar
System are obtained from the JPL ephemerides DE421; the rotation of
the Earth is provided by the interpolation table made public by the
International Earth Rotation Service (IERS) and the coordinates
associated with the ground stations are expected to be available.

The procedure to compute $\mathbf{u}^{*}$ is based on a modified
Newton's method known in the literature as \emph{differential
  corrections method}. Let us define
\[
B={\partial\boldsymbol{\xi}\over\partial\mathbf{u}}(\mathbf{u}),
\quad \quad C=B^TWB,
\] 
which are called the {\it design} matrix and
the {\it normal} matrix, respectively. Then the correction:
\begin{equation}\label{eq:2}
\Delta \mathbf{u}=C^{-1}D \quad \textrm{with   } D=-B^TW\boldsymbol{\xi}
\end{equation}
is applied iteratively until either $Q$ does not change meaningfully
from one iteration to the other or $\Delta \mathbf{u}$ becomes smaller
than a given tolerance.

\subsection{Simulators and correctors}\label{subsec:simcorr}

\begin{figure}
\includegraphics[width=7.5cm]{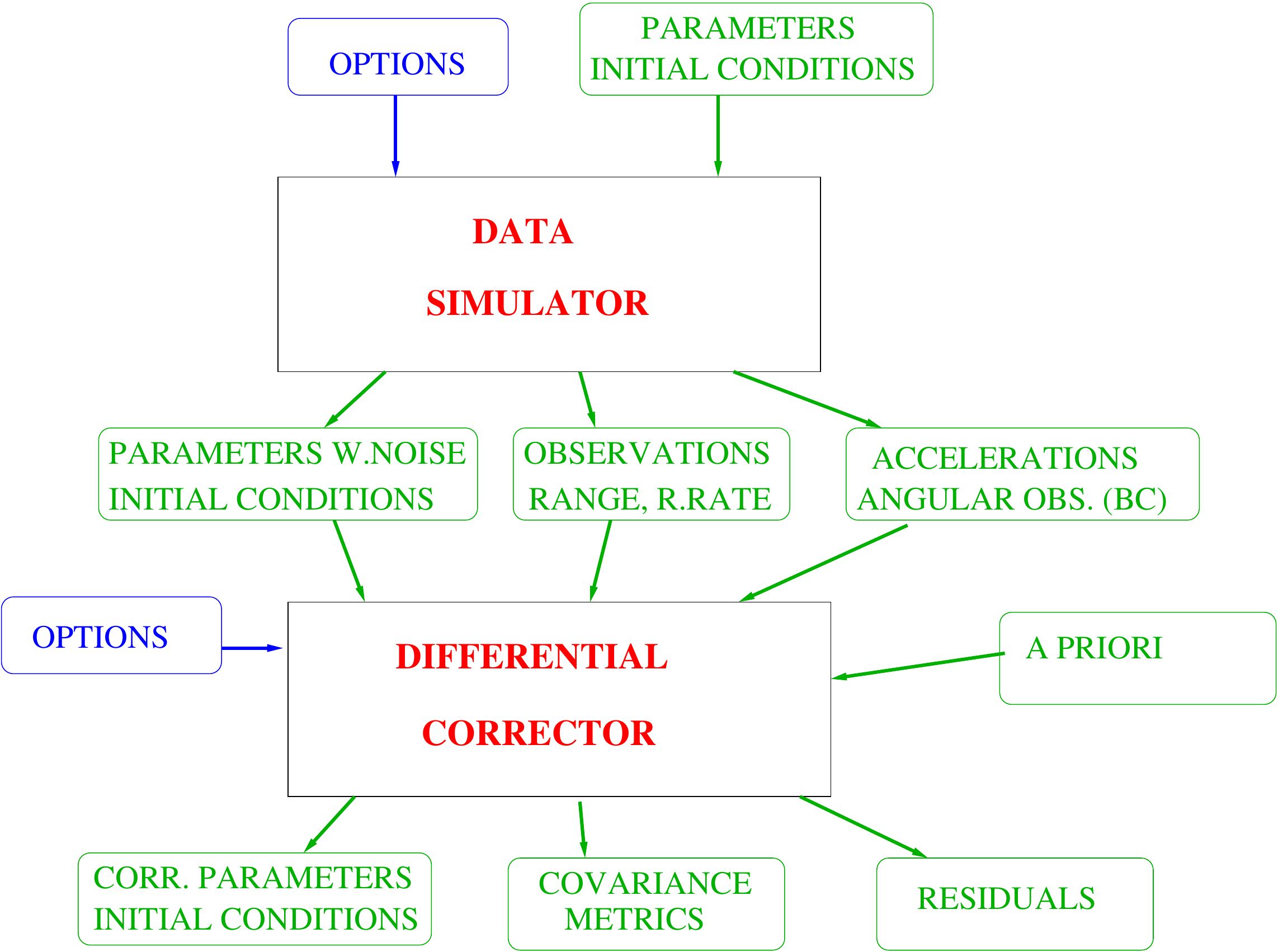}
\caption{The block diagram of a simple simulator
  setup. The black rectagles indicate the main programs, the
  rectangles with smoothed corners the data structures.}
\label{fig:simcor_blockdia}
\end{figure}

The structure of the overall software is outlined in
Fig.~\ref{fig:simcor_blockdia}. In general, developing a software
for OD and PE, the main programs belong to two categories: {\bf data
  simulators} (short: simulators) and {\bf differential correctors}
(short: correctors). The simulators generate simulated observables
(range and range-rate, accelerations) and preliminary orbital
elements. The correctors solve for all the parameters which can be
determined by a least squares fit (possibly constrained and/or
decomposed in a multi-arc structure). The simulators have a
fundamental role because the real data are available many years in the
future, while in the present we have the necessity of studying the
possibility to estimate some parameters in order to achieve the
scientific goals.

The program structure of the simulator is comparatively simple, with
all the complexity in the dynamics (Sec.~\ref{sec:dyn}), observations
(Sec.~\ref{sec:obs}) and error models. The corrector structure has to
be designed in a very careful way. One of the goals of our software
development is to be able to exploit parallel computing, especially
for the most computationally expensive portion of the processing.  The
propagation of the jovicentric orbit contains most of the
computational complexity, together with the light-time
computation. Thus the idea is to parallelize the computation of the
jovicentric dynamics and of the observables, with the relative partial
derivatives, and also the composition of the design matrix, the normal
matrix, and the inversion of the local normal matrices.

The choice of the method for the estimation of parameters, following
what we have done for the similar software for the BepiColombo RSE
(see, for example, \cite{ACMT}), should be essentially twofold:
\begin{itemize}
\item {\bf multi-arc strategy}: according to this method, every single arc
has its own set of initial conditions; in this way, the actual errors
in the orbit propagation, due to lack of knowledge in the
non-gravitational dynamical model, can be reduced by an
over-parametrization of the initial conditions;
\item {\bf constrained multi-arc strategy}: this method is 
established on the idea that each observed arc belongs to the same
object (the S/C) and thus the orbits corresponding to two subsequent
extended arcs (we call ``extended arc'' an observed arc broadened out
from half of the dark period before it to half of the dark period
after it) should coincide at the connection time in the middle of the
non-observed interval.
\end{itemize}

\noindent In the development of the corrector we are going to
investigate both approaches, even if we are quite convinced that the
multi-arc approach will be the right way.

\subsection{Propagators and multiple dynamics}

The observables depend on multiple dynamics, thus the main programs
need to have available the propagated state for each dynamics (for the
list of dynamics, see Sec.~\ref{sec:dyn}). This is obtained in
different ways, depending upon the dynamics.
\begin{itemize}
\item For the dynamics which have to be propagated by numerical
  integration that is the S/C around Jupiter, the Barycenter of Jovian
  System orbit, and the S/C interplanetary orbit we call a propagator
  which uses the corresponding dynamic module and solves the equation
  of motion, for the requested time interval. The states (time,
  position, velocity, acceleration) are stored in a memory stack, from
  which interpolation is possible with the required accuracy. Then,
  when the state is needed to compute the observables, the dynamics
  stacks are consulted and interpolated by the propagator modules.

\item The dynamics of the Jupiter rotation is represented by a
  semi-empirical model and the state is computed
  as an explicit function of time.

\item For the dynamics of the Earth rotation, for which an
  interpolation table is already available from an external source
  (IERS), the interpolation is performed inside the corresponding
  dynamical module. 

\item For the planetary ephemerides and the Jupiter's satellites
  ephemerides the interpolation tables are provided from JPL, together
  with the interpolation software which has been adapted for the
  insertion in our code.  The time ephemerides, to convert between
  different time systems, are not currently available from JPL and
  therefore they are propagated in a suitable dynamics.
\end{itemize}

There could be more than one propagator module, in particular because
of the need of quadruple precision computations for the heliocentric
orbits. However, the architecture is compatible with $M$
propagators/interpolators acting on $N$ dynamics.

\section{Dynamics}\label{sec:dyn}

In this section, for the case of Juno around Jupiter, the dynamics
which affect the observables are described. For the SCE, thinked to
constrain the PN parameter $\gamma$, a separate software has been
implemented, but it will not be described in this paper: for more
details, see \cite{TDMS}.

\subsection{Time ephemerides}

\begin{figure}
\begin{center}
\includegraphics[width=6cm]{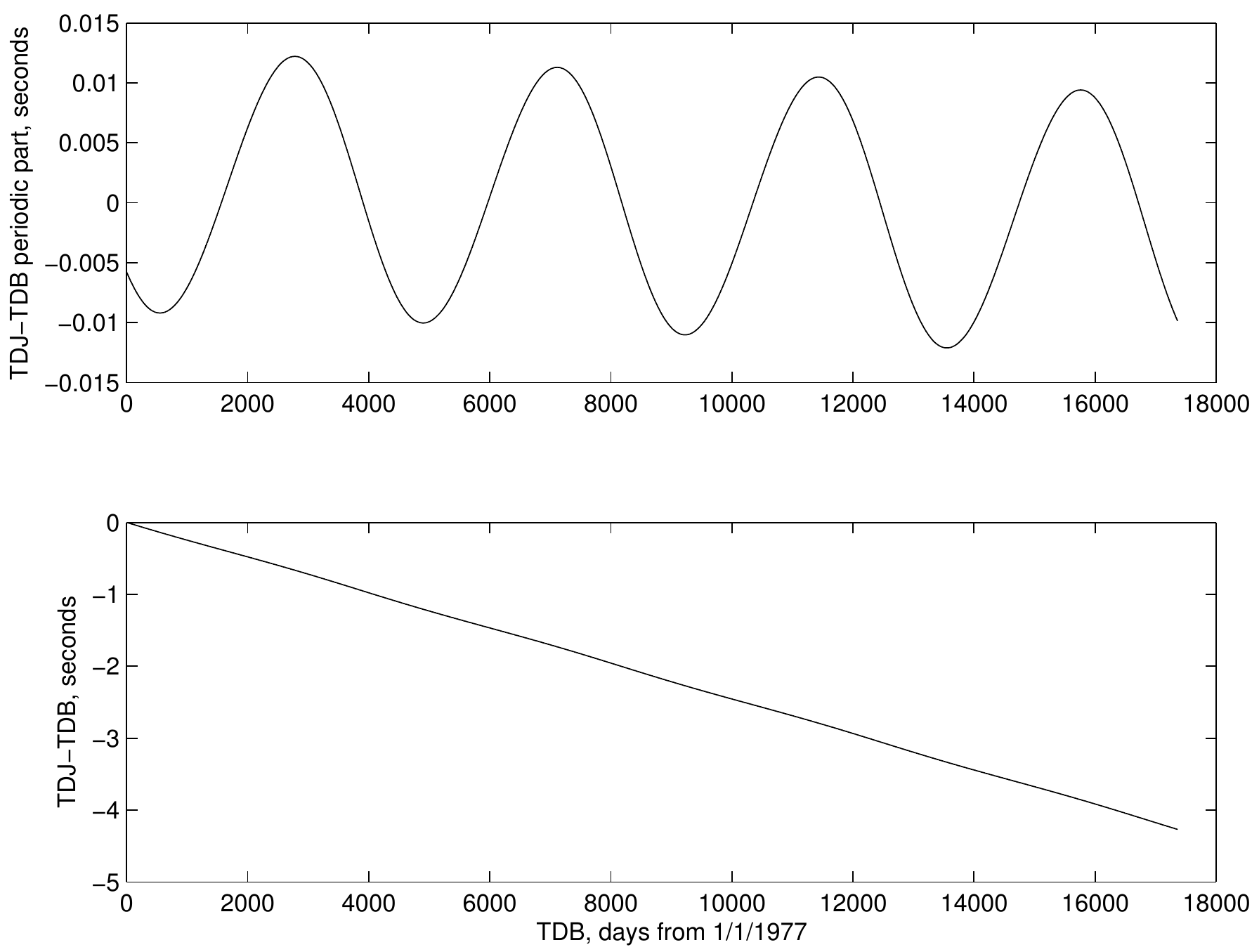}
\end{center}
\caption{Top: the oscillatory term, with the
period of Jupiter's orbit. Bottom: Linear drift with respect to TDB.}
\label{fig:tdj}
\end{figure}

The time coordinate which must be used for the formulation of the full
relativistic equations of motion for an orbiter around a planet
(\cite{DSX4}) is the dynamical time relative to the planet,
essentially the proper time of a body moving with the center of mass
of the planet (\cite{TMV}). For Jupiter we need to define a Jupiter
Dynamical Time (TDJ) containing terms of 1-PN order depending mostly
upon the distance $r_0$ from the Sun and velocity $v$ of Jupiter. The
relationship with the TDB scale, truncated to 1-PN order (we drop the
$O(c^{-4})$ terms on the right hand side, that are in principle known,
but certainly not needed for our purposes), is given by a differential
equation
\[
\dertot{t_\ros{TDJ}}{t_\ros{TDB}}= 1-\frac{v^2}{2\,c^2}- 
\sum_{k\neq jup}\, \frac{G\,m_k}{c^2\,r_{k}} \,\, , 
\]
which can be solved by a quadrature formula once the orbits of
Mercury, the Sun and the other planets are known. Figure~\ref{fig:tdj}
plots the output of such a computation, showing a drift due to the
non-zero average of the post-Newtonian term.  

The dynamical times relative to the planets, including the one
relative to the Earth, Terrestrial Dynamical Time (TDT, also shortly
indicated as TT) are defined as a function of the Barycentric
Dynamical Time (TDB) by the same separable differential equation. In our
software We use a Gaussian quadrature formula to generate an interpolation table 
for the conversion from TDB to TDT, TDJ. This table is precomputed by a
separate main program and it can be read by all other programs: a
suitable module uses the interpolation to compute all conversions of
time coordinates, that is it implements an internal system of time
ephemerides.

\subsection{S/C around Jupiter}

The orbit of Juno, when already in orbit around Jupiter, has an
equation of motion containing:

\begin{itemize}
\item the spherical harmonics of the gravity field of Jupiter,
\item the non gravitational perturbations, including
  direct radiation pressure, pressure from radiation reflected and
  emitted by Jupiter, thermal emission;
\item solar and planetary differential attractions and
  tidal perturbations,
\item Jupiter's satellite differential attractions and
  tidal perturbations,
\item relativistic corrections.
\end{itemize}

There are two main relativistic corrections in the jovicentric orbit:
one is due to the need to use proper time TDJ affected by the
gravitational potential at the center of mass of Jupiter; the second
is the Lense-Thirring (LT) effect depending upon the angular momentum
of Jupiter.

\subsection{Orbit of BJS}

The orbit of the BJS, centered at the Solar System Barycenter (SSB)
has an equation of motion (computed in TDB) including:
\begin{itemize}
\item the Newtonian attraction from the Sun and the planets;
\item the relativistic PPN corrections, including the
  PN parameters $\gamma, \beta$ and the Sun's dynamic oblateness;
\item the effects of possible violations of GR are limited to possible
  changes to the parameter $\gamma$, which shall be tested in a SCE in
  the cruise phase;
\item the effect of the satellites on the motion of the BJS.
\end{itemize}

The decomposition of the motion of Jupiter into the sum of two state
vectors, the BJS centered at the SSB and Jupiter centered at the BJS,
is classical in Newtonian mechanics (Jacobian vectors) but far from
trivial in a full relativistic PPN formalism.  Thus for the BJS the
attractions listed above need to be computed as the resultant
attraction on Jupiter and on the major satellites, but there may be
additional relativistic terms which are relevant.

The effect of the satellites on the orbit of the BJS is small
(Roy-Walker smallness parameter $1.5 \times 10^{-10}$), but large
enough to affect the residuals of the RSE at a level above the noise.

\subsection{Rotation of Jupiter and of the Earth}

For the rotation of Jupiter we use a semi-empirical model, containing
parameters defining the model of the Jupiter's rotation, essentially two angles
$\delta_1$ and $\delta_2$ specifying the direction of the rotation axis with respect to the
direction given by the IAU (\cite{IAUCART}).

For the Earth we are using the
interpolation tables made public by the International Earth Rotation
Service (IERS), because there is no way to solve for Earth rotation
parameters from observations at Jupiter (at accuracies competitive
with other available measurements),

The same argument applies to the station coordinates, which we have to
assume are supplied by the ground station with the required accuracy,
including corrections for the antenna motion.

\subsection{Planetary ephemerides}

The current state, as a function of time, of the planets Mercury to
Neptune (excluding Jupiter and considering also Pluto) are read from
the JPL ephemerides (currently the DE421 version, \cite{FWB}) as
Chebichev polynomials, which are interpolated with the JPL well tested
algorithm. For Jupiter satellites the ephemerides are provided by JPL
in the form of SPICE kernels (\cite{J}): the SPICE software has been
linked and suitable interfaces have been implemented in the code.
These ephemerides use the TDB time coordinate.

We need also to take into account asteroid perturbations on the orbit
of the BJS. The software to generate asteroid ephemerides
interpolation tables is available, and the interface has been built,
to be used with as many asteroids as needed. At the moment, for
consistency with DE421 we use the perturbations from 343 asteroids,
each one with a mass as assigned by the ephemerides.

\section{Observables}\label{sec:obs}

\begin{figure}
\includegraphics[width=6.5cm]{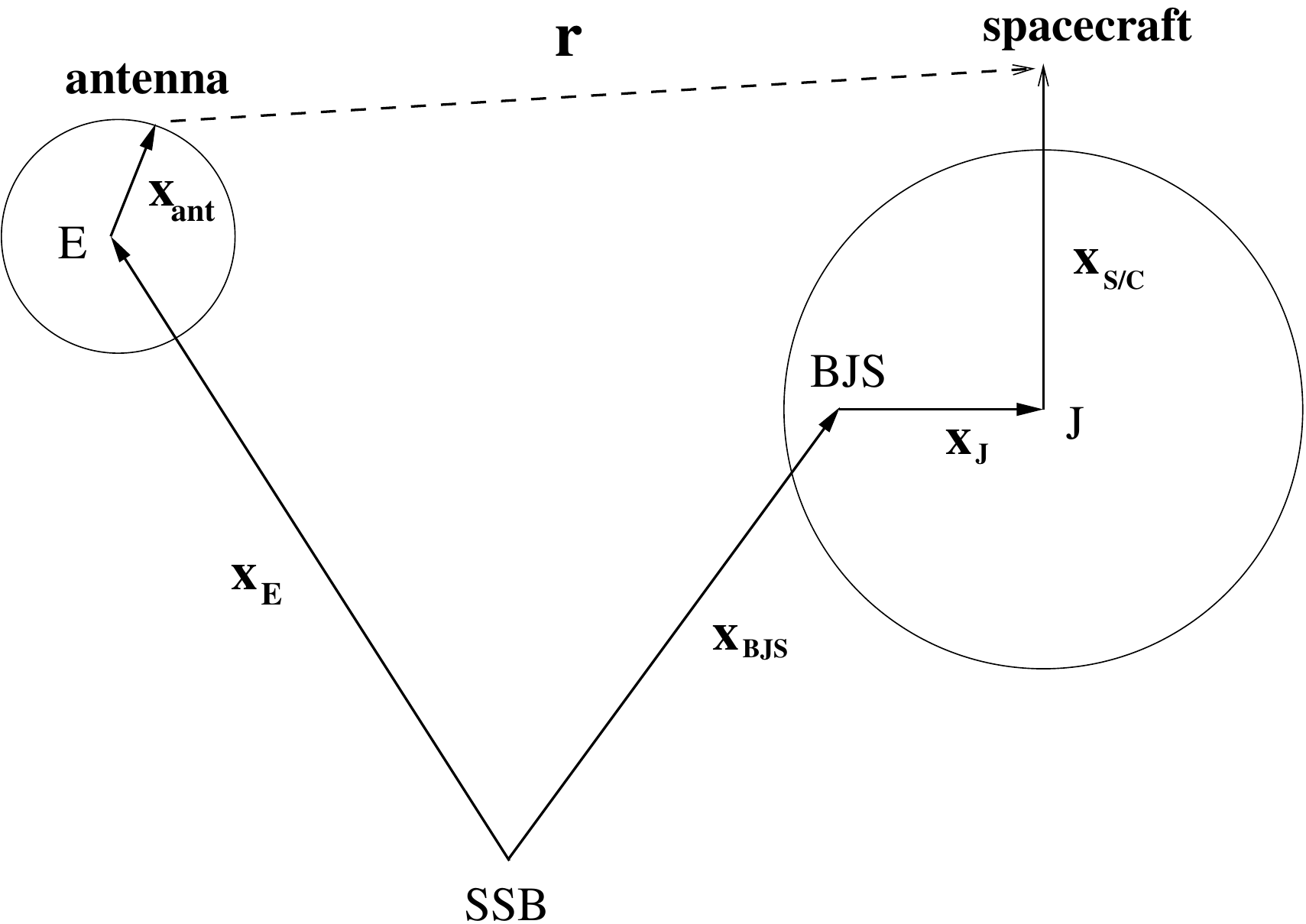}
\caption{Geometric sketch of the vectors involved in the computation
  of the range. SSB is the Solar System Barycenter, J is the center of
  Jupiter, BJS is the barycenter of the jovian system, E is the center
  of the Earth.}
\label{fig:range}
\end{figure}

As it is well clear in space navigation (see, for instance,
\cite{MOY}), the distance to a spacecraft cannot be computed by an
explicit analytic formula from the state of the ground station and the
spacecraft at the same time, but we need some {\em light-time
  iterations}.  In the Juno case the equation giving the range
(distance antenna-spacecraft, equivalent to the light-time up to a
factor $c$, see Fig.\ref{fig:range}) is
\begin{equation}
r=|(\x_{S/C}+\x_{BJS}+\x_J)-(\x_E+\x_{ant})|+S(\gamma)
\label{eq:juno_range}
\end{equation}
where $\x_{S/C}$ is the jovicentric position of the orbiter,
$\x_{BJS}$ is the position of the center of mass of Jupiter's system
in a reference system with origin at SSB, $\x_J$ is the vector from
the BJS to the center of mass of Jupiter, $\x_E$ is the position of
the Earth center of mass with respect to SSB, $\x_{ant}$ is the
position of the reference point of the ground antenna with respect to
the center of mass of the Earth, and $S(\gamma)$ is the \emph{Shapiro
  effect}. 

Using (\ref{eq:juno_range}) means to model the space as a
flat arena ($r$ is an Euclidean distance) and the time as an absolute
parameter.  This is obviously not possible because it is clear that,
beyond some threshold of accuracy, space and time have to be
formulated within the framework of Einstein's theory of gravity
(GR). Moreover we have to take into account the different times at
which the events have to be computed: the transmission of the signal
at the transmit time ($t_t$), the signal at the Jupiter orbiter at the
time of bounce ($t_b$) and the reception of the signal at the receive
time ($t_r$). Formula (\ref{eq:juno_range}) is used as a starting
point to construct a correct relativistic formulation; with the word
``correct'' we do not mean all the possible relativistic effects, but
the effects that can be measured by the experiment.

The five vectors involved in formula (\ref{eq:juno_range}) have to be
computed at their own time, the epoch of different events: e.g., ${\bf
  x}_{\rm ant}$ and ${\bf x}_{\rm E}$ are computed at both the antenna
\emph{transmit time} $t_t$ and \emph{receive time} $t_r$ of the
signal, ${\bf x}_{\rm S/C}$, ${\bf x}_{\rm BJS}$ and ${\bf x}_{\rm J}$
are computed at the \emph{bounce time} $t_b$ (when the signal has
arrived to the orbiter and is sent back, with correction for the delay
of the transponder). In order to be able to perform the vector sums
and differences, these vectors have to be converted to a common
space-time reference system, the only possible choice being some
realization of the BCRS (Barycentric Celestial Reference System). We
adopt for now a realization of the BCRS that we call SSB reference
frame and in which the time is a re-definition of the TDB; other
possible choices, such as TCB (Barycentric Coordinate Time), only can
differ by linear scaling. The TDB choice of the SSB time scale entails
also the appropriate linear scaling of space-coordinates and planetary
masses as described for instance in \cite{kli10}.

The vectors ${\bf x}_{\rm BJS}$ and ${\bf x}_{\rm E}$ are already in
SSB as provided by numerical integration and external ephemerides;
thus the vectors ${\bf x}_{\rm ant}$, ${\bf x}_{\rm J}$ and ${\bf
  x}_{\rm S/C}$ have to be converted to SSB from the geocentric and
jovicentric systems, respectively. Of course the conversion of
reference system implies also the conversion of the time coordinate.
There are three different time coordinates to be considered. The
currently published planetary ephemerides are provided in TDB.  The
observations are based on averages of clock and frequency measurements
on the Earth surface and the time coordinate is TT. Thus for each
observation the times of transmission $t_t$ and reception $t_r$ need
to be converted from TT to TDB to find the corresponding positions of
the planets by combining information from the pre-computed ephemerides
and the output of the numerical integration for the BJS. This time
conversion step is necessary for the accurate processing of each set
of interplanetary tracking data; the main term in the difference
TT-TDB is periodic, with period 1 year and amplitude $\simeq 1.6\times
10^{-3}$ s, while there is essentially no linear trend, as a result of
a suitable definition of the TDB.

From now on, in accordance with \cite{kli10}, we shall call the
quantities related to the SSB frame ``TDB-compatible'', the quantities
related to the geocentric frame ``TT-compatible'', the quantities
related to the jovicentric frame ``TDJ-compatible'' and label them TB,
TT and TJ, respectively.

The differential equation giving the local time $T$ as a function of the 
SSB time $t$ , which we are currently assuming to be TDB, is the following:
\begin{equation}
\dertot{T}{t}= 1- \frac{1}{c^2}\;\left[U+ \frac{v^2}{2}-L\right]\,\,,
\label{eq:dert}
\end{equation}
where $U$ is the gravitational potential (the list of contributing
bodies depends upon the accuracy required: in our implementation we
use Sun, Mercury to Neptune, Moon) at the planet center and $v$ is the
SSB velocity of the same planet. The constant term $L$ is used to
perform the conventional rescaling motivated by removal of secular
terms.

The space-time transformations to perform involve essentially the
position of the antenna and the position of the orbiter.  The
geocentric coordinates of the antenna should be transformed into
TDB-compatible coordinates; the transformation is expressed by the
formula
 \[
 {\bf x}_{\rm ant}^{TB}={\bf x}_{\rm ant}^{TT}\,\left
 (1-\frac{U}{c^2}-L_C \right)- \frac{1}{2}\,\left ( \frac{{\bf v}_{\rm
 E}^{TB}\cdot {\bf x}_{\rm ant}^{TT}}{c^2} \right )\, {\bf v}_{\rm
 E}^{TB}\,\,,
 \]
where $U$ is the gravitational potential at the geocenter (excluding
the Earth mass), $L_C=1.48082686741\, \times \,10^{-8} $ is a scaling
factor given as definition, supposed to be a good approximation for
removing secular terms from the transformation and ${\bf v}_{\rm
E}^{TB}$ is the barycentric velocity of the Earth. The next formula
contains the effect on the velocities of the time coordinate change,
which should be consistently used together with the coordinate change:
 \[
 {\bf v}_{\rm ant}^{TB} = \left[{\bf v}_{\rm ant}^{TT}\,\left
 (1-\frac{U}{c^2}-L_C \right)- \frac{1}{2}\,\left ( \frac{{\bf v}_{\rm
 E}^{TB}\cdot {\bf v}_{\rm ant}^{TT}}{c^2} \right )\, {\bf v}_{\rm
 E}^{TB}\right] \,\left[\dertot Tt\right]\,\,.
 \]
 Note that the previous formula contains the factor $dT/dt$ (expressed
 by (\ref{eq:dert})) that deals with a time transformation: $T$ is
 the local time for Earth, that is TT, and $t$ is the corresponding
 TDB time.

 The jovicentric coordinates of the orbiter have to be transformed
 into TDB-compatible coordinates through the formula
 \[
 {\bf x}_{\rm S/C}^{TB}={\bf x}_{\rm S/C}^{TJ}\,\left (1-\frac{U}{c^2}-L_{CJ}
 \right)- \frac{1}{2}\,\left ( \frac{{\bf v}_{\rm M}^{TB} \cdot {\bf
 x}_{\rm S/C}^{TJ}}{c^2} \right )\, {\bf v}_{\rm M}^{TB}\,\,,
 \]
 where $U$ is the gravitational potential at the center of mass of
 Jupiter (excluding the Jupiter mass) and $L_{CJ}$ could be used to
 remove the secular term in the time transformation (thus defining a
 TJ scale, implying a rescaling of the mass of Jupiter). We believe
 this is not necessary: the secular drift of TDJ with respect to other
 time scales is significant, but a simple iterative scheme is very
 efficient in providing the inverse time transformation.  Thus we set
 $L_{CJ}=0$, assuming the reference frame is TDJ-compatible.
As for the antenna we have a formula expressing the velocity
transformation that contains the derivative of time $T$ for Jupiter,
that is TDJ, with respect to time $t$, that is TDB:
 \begin{eqnarray*}
 {\bf v}_{\rm S/C}^{TB} & = & \left[ {\bf v}_{\rm S/C}^{TJ}\,\left
 (1-\frac{U}{c^2}-L_{CJ} \right)-\frac{1}{2}\,\left ( \frac{{\bf
 v}_{\rm M}^{TB}\cdot {\bf v}_{\rm S/C}^{TJ}}{c^2} \right )\,{\bf
 v}_{\rm M}^{TB} \right ] \cdot \\
 & & \,\left[\dertot{T}{t} \right]
 \end{eqnarray*}

\noindent For these coordinate changes, in every formula we neglected
the terms of the SSB acceleration of the planet center, because they
contain beside $1/c^2$ the additional small parameter (distance from
planet center)/(planet distance to the Sun), which is of the order of
$10^{-4}$.

The correct modeling of space-time transformations is not sufficient
to have a precise computation of the signal delay: we have to take
into account the general relativistic contribution to the time delay
due to the space-time curvature under the effect of the Sun's
gravitational field (or Jupiter gravitational field), the {\em Shapiro
  effect} \cite{SHAP}. The Shapiro time delay $\Delta t$ at the 1-PN
level, according to \cite{MOY}, is
\[
\Delta t=\frac{(1+\gamma)\,\mu_0}{c^3}\, \ln \, \left (
\frac{r_t+r_r+r}{r_t+r_r-r} \right )\,, \quad 
S(\gamma)=c\,\Delta t 
\]
where $r_t=|{\bf r}_{\rm t}|$ and $r_r=|{\bf r}_{\rm r}|$ are the
heliocentric distances of the transmitter and the receiver at the
corresponding time instants of photon transmission and reception,
$\mu_0$ is the gravitational mass of the Sun (or Jupiter) and $r=|{\bf
  r}_r-{\bf r}_t|$. Parameter $\gamma$ is the only post-Newtonian
parameter used for the light-time effect and, in fact, it could be
best constrained during superior conjunction.
The question arises whether the very high signal to noise in the range
requires other terms in the solar gravity influence, due to
higher-order corrections when the radio waves are passing near the
Sun, at just a few solar radii (and thus the denominator in the
log-function of the Shapiro formula is small). These corrections 
are of order $2$, (containing a factor $1/c^4$): the relevant
correction is most easily obtained by adding $1/c^4$ terms in the
Shapiro formula, due to the bending of the light path:
\[
S(\gamma)=\frac{(1+\gamma)\,\mu_0}{c^2}\, \ln
\,\left(\frac{r_t+r_r+r+\frac{(1+\gamma)\,\mu_0}{c^2}}
{r_t+r_r-r+\frac{(1+\gamma)\,\mu_0}{c^2}}\right) \,\,.
\]

\begin{figure}
\includegraphics[width=7cm]{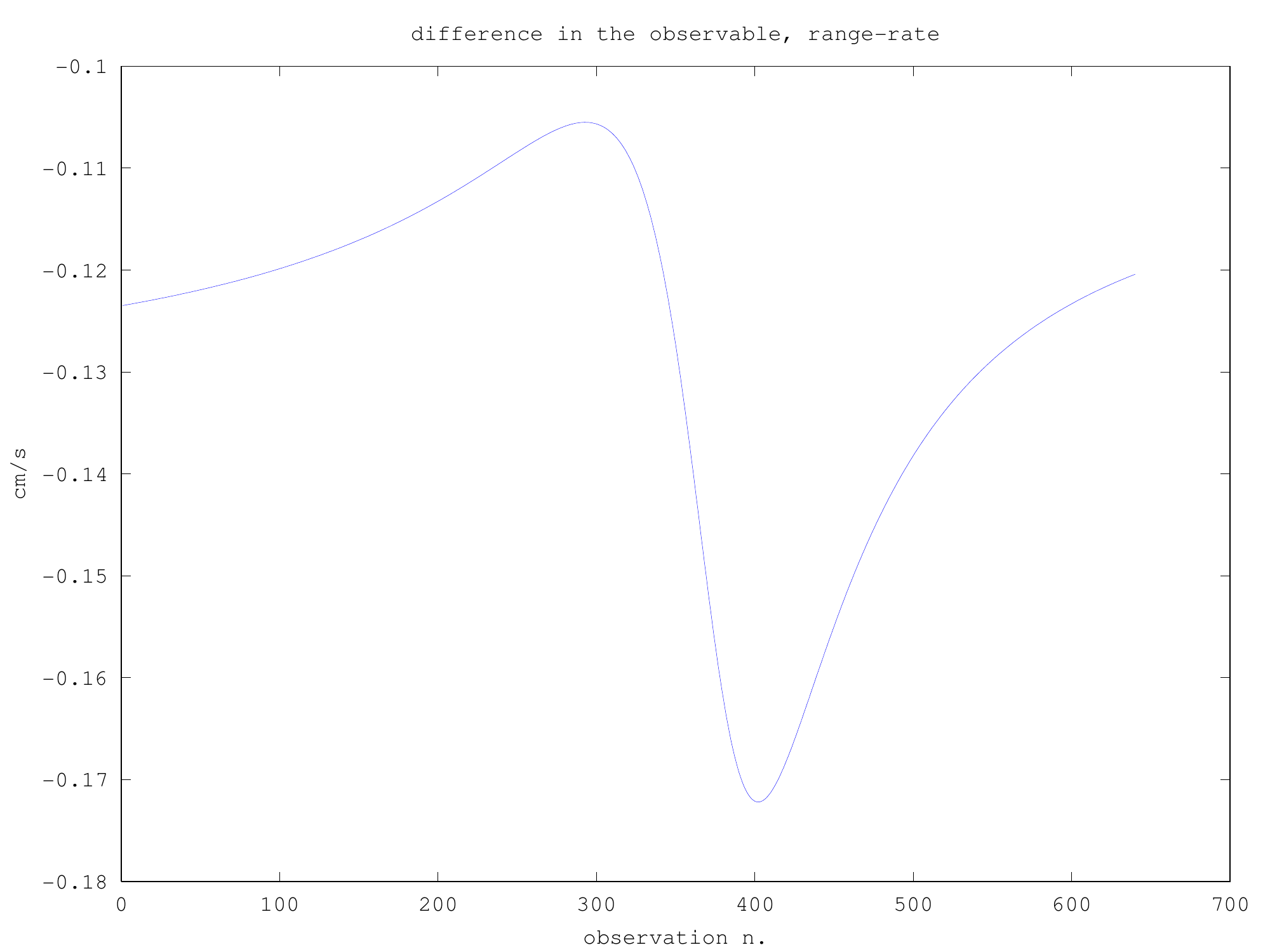}
\caption{Total amount of the Shapiro effect, both for the Sun and Jupiter, in range-rate observable
over 2-year simulation.}
\label{fig:shap}
\end{figure}

This formulation has been proposed by \cite{MOY} and it has been
justified in the small impact parameter regime by much more
theoretically rooted derivations by other
authors.  The total amount of the
Shapiro effect in range-rate is shown in Figure \ref{fig:shap}: see \cite{TMV} for 
a detailed description of the Shapiro delay for the range-rate observable.

Of course, the light-time computation is an iterative procedure. In
particular, since radar measurements are usually referred to the
receive time $t_r$, the observables are seen as functions of this
time, and the computation sequence works backward in time: starting
from $t_r$, the bounce time $t_b$ is computed iteratively (down-leg
iterations), and, using this information the transmit time $t_t$ is
computed (up-leg iterations).

\subsection{Integrated observables}

A problem well known to radio science experts is that for top
accuracy the range-rate measurement cannot be the instantaneous value (up: up-leg, do: down-leg)
\[
\dot r(t_r)=(\dot r_{up}(t_r)+\dot r_{do}(t_r))/2 \,\, .
\]  
In fact, the measurement is not instantaneous: an accurate measure of a Doppler
effect requires to fit the difference of phase between carrier waves,
the one generated at the station and the one returned from space,
accumulated over some {\em integration time} $\Delta$, typically
between $10$ and $1000$ s. Thus the observable is really a difference
of ranges
\[
r(t_b+\Delta/2)-r(t_b-\Delta/2)=\int_{t_b-\Delta/2}^{t_b+\Delta/2}\;
\dot r(s)\; ds
\]
or, equivalently, an averaged value of range-rate over the integration interval
\begin{equation}
\dot r_{\Delta}(t_r)=\frac 1\Delta\; \int_{t_b-\Delta/2}^{t_b+\Delta/2}\;
\dot r(s)\; ds\ .
\label{eq:rrate_mode3}
\end{equation}

We decided to implement a quadrature formula to approximate the
integral of eq.~(\ref{eq:rrate_mode3}) because it allows to control
much better the rounding off problems, because the relative accuracy
in range-rate is by no means as extreme as the one required to
implement the $r(t_b+\Delta/2)-r(t_b-\Delta/2)$ computation. 
In \cite{MTVLC} some figures describing this issue are present.

\section{Sensitivity analysis}\label{sec:sim}

In this section we want to discuss some results concerning the
scientific goals of the mission obtained with simulated data and
running \emph{sensitivity tests}. This kind of tests consists in
varying of some extent the default value of a parameter and to see if
the observable is affected, in the limit of the accuracy, by such
variation. But the first thing to do is to compile a list of all the
effects, both dynamical (that is acting on the orbit) and
observational (that is acting on the light-time and its time
derivative) that can change the motion of the spacecraft in orbit
around Jupiter and then modify the observable down to the
observational noise level. The major dynamical effects are described
in Tab.~\ref{tab:pert}: the table contains the effect (e.g. the
Lense-Thirring effect), its formulation and its value in cm/s$^2$. 

We can note that, a part from the gravitational field of Jupiter, the
strongest perturbation is the gravitational one of the Io satellite,
while the gravitational perturbations from Europa and Ganymede are of
the same order of the solid tides due to Io and the relativistic
perturbation of Jupiter. It is clear from the table that the inclusion
of Galilean satellites in the model is necessary, but, in our
investigations, we also found that the satellites Amalthea and Thebe
play an important role, as we will show in
Subsec.~\ref{subsec:sat}. At the moment we are not taking into account 
non-gravitational effects, although we are aware that to get a good determination of 
the parameters of interest it will be necessary to estimate these effects.

In the next subsections we are going to
display some figures showing the changes in the observable range and
range-rate, even if, for the Juno mission, only the doppler data,
corresponding to range rate, will be available for accurate radio science experiments.
We want to stress that this preliminary work is necessary but not sufficient to know if a parameter estimation
will experience a success, there a lot of sources of error that could be taken into
account, but this aspect will be the subject of a next paper.

\begin{center}
\begin{table*}
{\small
\hfill{}
\begin{tabular}{l c c l}
  \hline Cause & Formula & Parameters& Value cm/s$^2$ \\ 
    \hline {\bf Jupiter monopole} & 
  ${GM_{jup}}/{r^2}=F_0$ & $GM_{jup}$ 
  & $2.3 \times 10^3$\\ 
  {\bf Jupiter oblateness}& $3\,F_0\,\overline
  C_{20}\,{R_{jup}^2}/{r^2}$ &
  $\overline C_{20}$ & $1.2 \times 10^{2}$\\
  {\bf Jupiter $C_{40}$}&$5\,F_0\,\overline
  C_{40}\,{R_{jup}^4}/{r^4}$ &
  $\overline C_{40}$ & $5.0 $\\
  {\bf Jupiter $C_{60}$}&$7\,F_0\,\overline
  C_{60}\,{R_{jup}^6}/{r^6}$ &
  $\overline C_{60}$ & $-3.54 \times 10^{-1} $\\
  {\bf Jupiter $C_{30}$}&$4\,F_0\,\overline
  C_{30}\,{R_{jup}^3}/{r^3}$ &
  $\overline C_{30}$ & $4.7 \times 10^{-3}$\\
  {\bf Jupiter $S_{22}$}&$3\,F_0\,\overline
  S_{22}\,{R_{jup}^2}/{r^2}$ &
  $\overline S_{22}$ & $-7.6 \times 10^{-5}$\\
  {\bf Jupiter triaxiality}&$3\,F_0\,\overline
  C_{22}\,{R_{jup}^2}/{r^2}$ &
  $\overline C_{22}$ & $3.52\times 10^{-5}$\\
  {\bf Io pert.} &
  $2\,{GM_{io}\,r}/{r_{j\,io}^3}$ & 
  $GM_{io}$ & $1.2 \times 10^{-3}$\\
  {\bf Europa pert.} &
  $2\,{GM_{eur}\,r}/{r_{j\,e}^3}$ & 
  $GM_{eur}$ & $1.6 \times 10^{-4}$\\
  {\bf Ganymede pert.} &
  $2\,{GM_{gan}\,r}/{r_{j\,g}^3}$ & 
  $GM_{gan}$ & $1.2 \times 10^{-4}$\\
  {\bf Relativistic Jupiter} &
  $F_0\,4\,G\,M_{jup}/(c^2\, r)$ & $G\,M_{jup}$ & 
  $1.7\times 10^{-4}$\\

  {\bf Solid tide (Io)} & $3k_2\,GM_{io}\,
  R_{jup}^5/(r_{J\,io}^3\,r^4)$ & $k_2$ & 
  $8.9 \times 10^{-4}$  \\
  {\bf Solid tide (Europa)} & $3k_2\,GM_{eur}\,
  R_{jup}^5/(r_{J\,eur}^3\,r^4)$ & $k_2$ & $1.2 
  \times 10^{-4}$  \\
  {\bf Solid tide (Io)} & $4k_3\,GM_{io}\,
  R_{jup}^7/(r_{J\,io}^4\,r^5)$ & $k_3$ & 
  $1.8 \times 10^{-4}$  \\
  {\bf Solid tide (Ganymede)} & $3k_2\,GM_{gan}\,
  R_{jup}^5/(r_{J\,gan}^3\,r^4)$ & $k_2$ & $
  8.2 \times 10^{-5}$  \\
  {\bf Solid tide (Io)} & $5k_4\,GM_{io}\,
  R_{jup}^9/(r_{J\,io}^5\,r^6)$ & $k_4$ & 
  $3.6 \times 10^{-5}$  \\
  {\bf Lense-Thir.} & $6\,GJ_{jup}|\dot{\mathbf r}|/{c^2}\,{r^3}$ & 
  $GJ_{jup}$ &
  $2.5 \times 10^{-5}$\\ 
  {\bf Callisto pert.} &
  $2\,{GM_{cal}\,r}/{r_{j\,c}^3}$ & 
  $GM_{cal}$ & $1.6 \times 10^{-5}$\\
  {\bf Solid tide (Europa)} & $4k_3\,GM_{eur}\,
  R_{jup}^7/(r_{J\,eur}^4\,r^5)$ & $k_3$ & $1.5 
  \times 10^{-5}$  \\
  {\bf Solid tide (Callisto)} & $3k_2\,GM_{cal}\,
  R_{jup}^5/(r_{J\,cal}^3\,r^4)$ & $k_2$ & $
  1.2 \times 10^{-5}$  \\

  {\bf Solid tide (Ganymede)} & $4k_3\,GM_{gan}\,
  R_{jup}^7/(r_{J\,gan}^4\,r^5)$ & $k_3$ & $
  6.5 \times 10^{-6}$  \\
  {\bf Sun pert.} & $2\,{GM_{ \odot}\,r}/{r_{\odot}^3}$ & 
  $GM_{\odot}$ &
  $4.7 \times 10^{-6}$\\
  {\bf Solid tide (Sun)} & $3k_2\,GM_{\odot}\,
  R_{jup}^5/(r_{J\,\odot}^3\,r^4)$ & $k_2$ & $
  3.1 \times 10^{-6}$  \\
  {\bf Solid tide (Europa)} & $5k_4\,GM_{eur}\,
  R_{jup}^9/(r_{J\,eur}^5\,r^6)$ & $k_4$ & $1.9 
  \times 10^{-6}$  \\
  {\bf Solid tide (Callisto)} & $4k_3\,GM_{cal}\,
  R_{jup}^7/(r_{J\,cal}^4\,r^5)$ & $k_3$ & $
  5.3 \times 10^{-7}$  \\
  {\bf Solid tide (Ganymede)} & $5k_4\,GM_{gan}\,
  R_{jup}^9/(r_{J\,gan}^5\,r^6)$ & $k_4$ & $
  4.8 \times 10^{-7}$  \\
  {\bf Amalthea pert.} &
  $2\,{GM_{ama}\,r}/{r_{J\,ama}^3}$ & 
  $GM_{ama}$ & $3.5 \times 10^{-7}$\\ 
  {\bf Solid tide (Amalthea)} & $3k_2\,GM_{ama}\,
  R_{jup}^5/(r_{J\,ama}^3\,r^4)$ & $k_2$ & $
  2.6 \times 10^{-7}$  \\
  {\bf Thebe pert.} &
  $2\,{GM_{the}\,r}/{r_{J\,the}^3}$ & 
  $GM_{the}$ & $1.4 \times 10^{-7}$\\ 
  {\bf Solid tide (Amalthea)} & $4k_3\,GM_{ama}\,
  R_{jup}^7/(r_{J\,ama}^4\,r^5)$ & $k_3$ & $
  1.2 \times 10^{-7}$  \\
  {\bf Solid tide (Thebe)} & $3k_2\,GM_{the}\,
  R_{jup}^5/(r_{J\,the}^3\,r^4)$ & $k_2$ & $
  1.0 \times 10^{-7}$  \\
  {\bf Solid tide (Amalthea)} & $5k_4\,GM_{ama}\,
  R_{jup}^9/(r_{J\,ama}^5\,r^6)$ & $k_4$ & $
  5.5 \times 10^{-8}$  \\
  {\bf Solid tide (Thebe)} & $4k_3\,GM_{the}\,
  R_{jup}^7/(r_{J\,the}^4\,r^5)$ & $k_3$ & $
  4.0 \times 10^{-8}$  \\
  {\bf Solid tide (Callisto)} & $5k_4\,GM_{cal}\,
  R_{jup}^9/(r_{J\,cal}^5\,r^6)$ & $k_4$ & $
  2.3 \times 10^{-8}$  \\
  {\bf Solid tide (Thebe)} & $5k_4\,GM_{the}\,
  R_{jup}^9/(r_{J\,the}^5\,r^6)$ & $k_4$ & $
  1.5 \times 10^{-8}$  \\
  {\bf Solid tide (Sun)} & $4k_3\,GM_{\odot}\,
  R_{jup}^7/(r_{J\,\odot}^4\,r^5)$ & $k_3$ & $
  3.4 \times 10^{-10}$  \\
  {\bf Solid tide (Sun)} & $5k_4\,GM_{\odot}\,
  R_{jup}^9/(r_{J\,\odot}^5\,r^6)$ & $k_4$ & $
  3.6 \times 10^{-14}$  \\
  \hline \hline
\end{tabular}}
\hfill{}
\caption{Accelerations acting on a spacecraft in orbit
    around Jupiter, in a planetocentric reference frame, with
    $r=75000$ km.}
\label{tab:pert}
\end{table*}
\end{center}

\subsection{Satellites and solid tides}\label{subsec:sat}

\begin{figure}
\includegraphics[width=7.5cm]{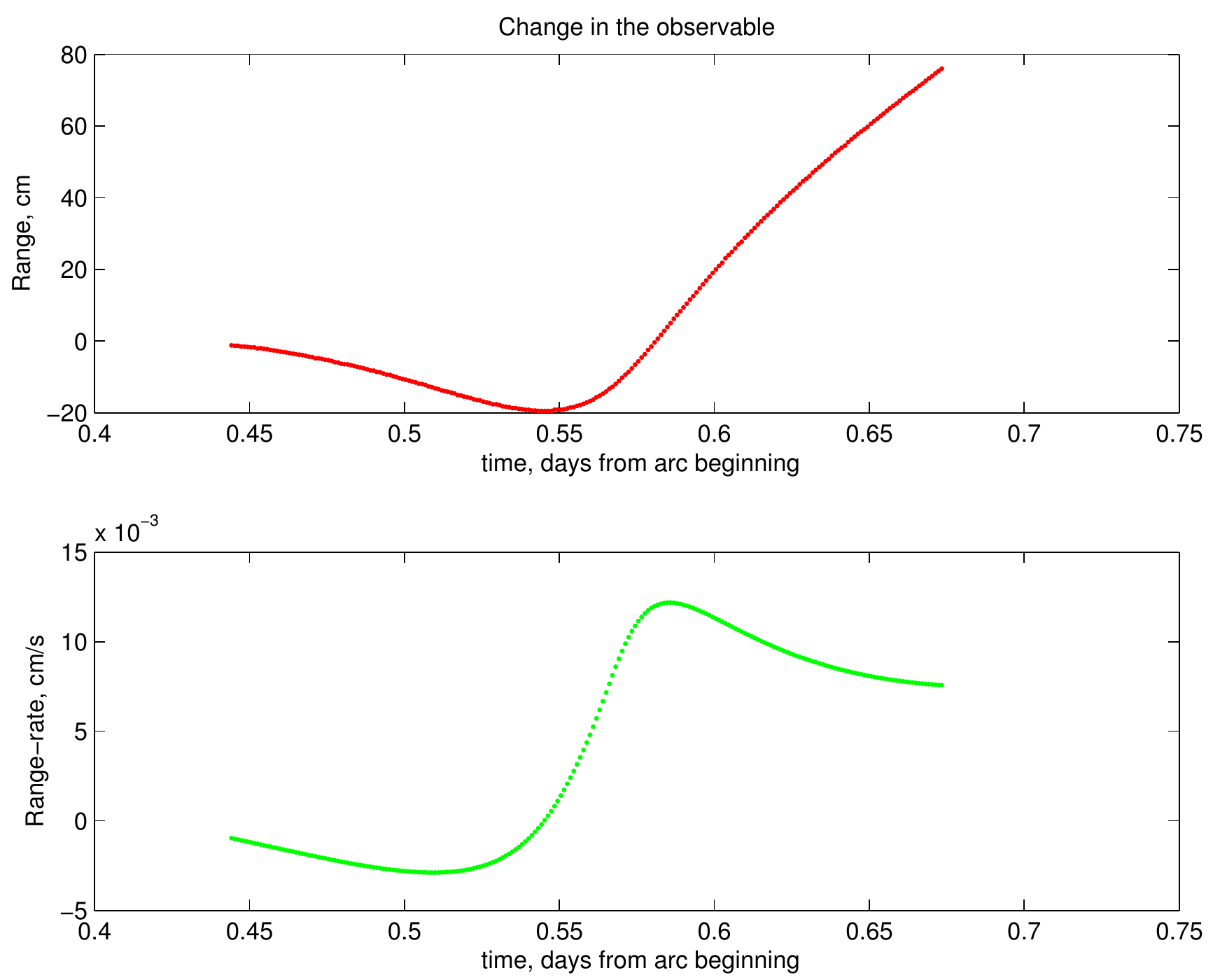}
\caption{Perturbations from Amalthea and Thebe.}
\label{fig:jsat4_6}
\end{figure}

As mentioned before, the role of the Galilean satellites is
fundamental, but also Amalthea and Thebe affects the observables. In
Fig.~\ref{fig:jsat4_6} we show the sensitivity of the Juno observables
(above range, below range-rate) to the perturbations from Amalthea and
Thebe, over an arc centered at perijove supposing a noise from KAT at
$\simeq 3\times 10^{-4}$ cm/s for an integration time of $1000$
seconds. The figure is obtained as difference between a run of the
software including in the model Amalthea and Thebe and a run with only
the Galilean satellites. The Signal-to-Noise ($S/N$) ratio is, for the
range-rate, about $50$, that is the signal from the two satellites is
distinguishable and we could try to improve our understanding on this
two satellites.

From Tab.~\ref{tab:pert} it is clear how important it is to take into
account the solid tides, not only due to the Sun but also those
resulting from the Galilean satellites: the next three figures show
how different degrees, described using the Love numbers $k_j$ as parameters to be determined, 
affect the observables.

\begin{figure}
\includegraphics[width=7.5cm]{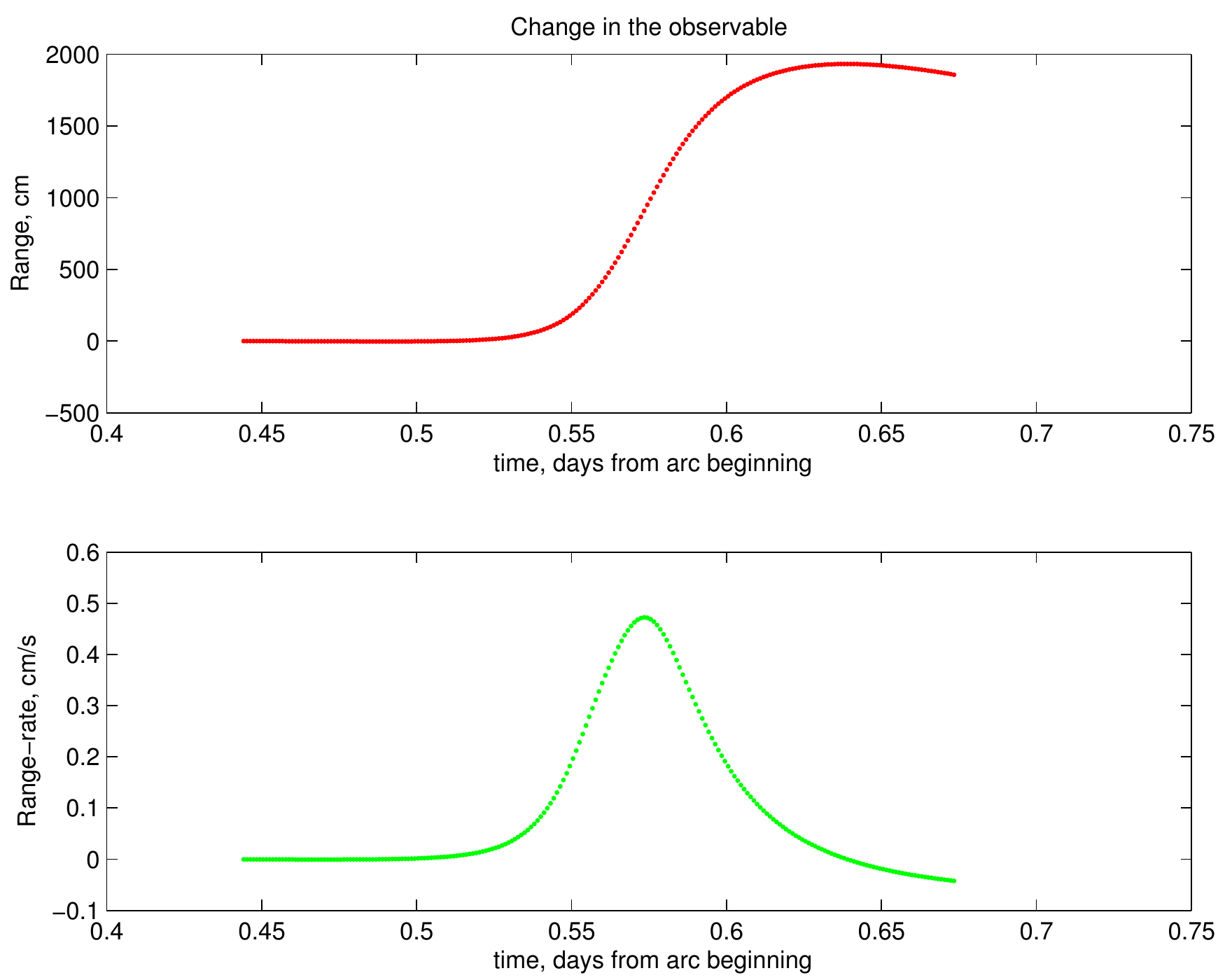}
\caption{Perturbations from the tidal degree 2 term from
all the Galilean satellites and the Sun.}
\label{fig:k2}
\end{figure}

Fig.~\ref{fig:k2} represents the perturbations from the tidal degree $2$
term from all the Galilean satellites and the Sun, acting together
with a frequency-independent $k_2$ (for the simulation $k_2=0.7$ has
been assumed) and it points out that the parameter $k_2$ could be
determined very well ($S/N \simeq 2000$).

\begin{figure}
\includegraphics[width=7.5cm]{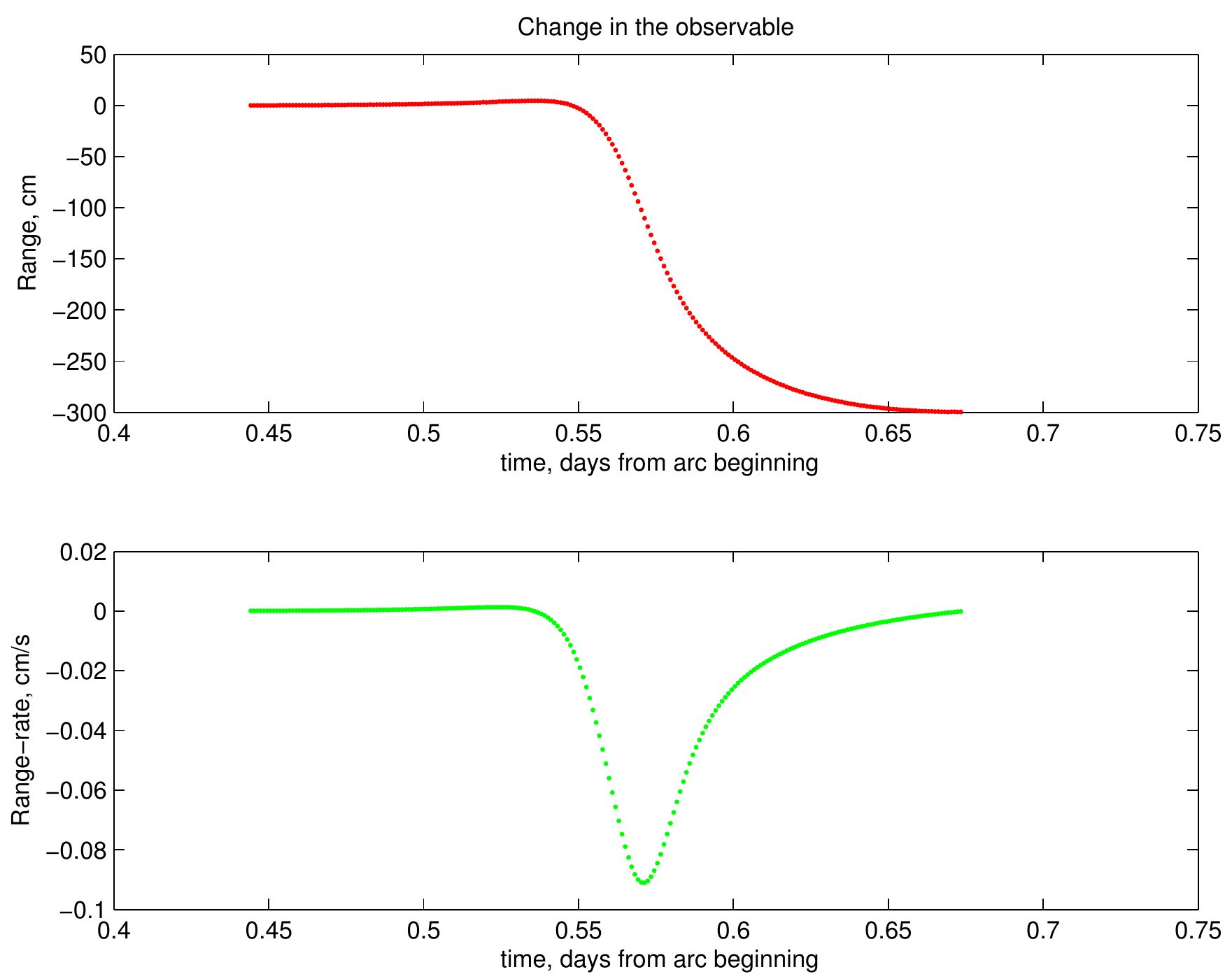}
\caption{Perturbation from the tidal degree 3 term from
all the Galilean satellites and the Sun.}
\label{fig:k3}
\end{figure}

Fig.~\ref{fig:k3} shows the sensitivity of the observables to the
tidal degree $3$ term from all the Galilean satellites and the Sun,
acting together with a frequency-independent $k_3$ (for the simulation
$k_3=0.7$ has been assumed): the $S/N$ is about $300$ and it means
that we could extract information about $k_3$ from the data.

\begin{figure}
\includegraphics[width=7.5cm]{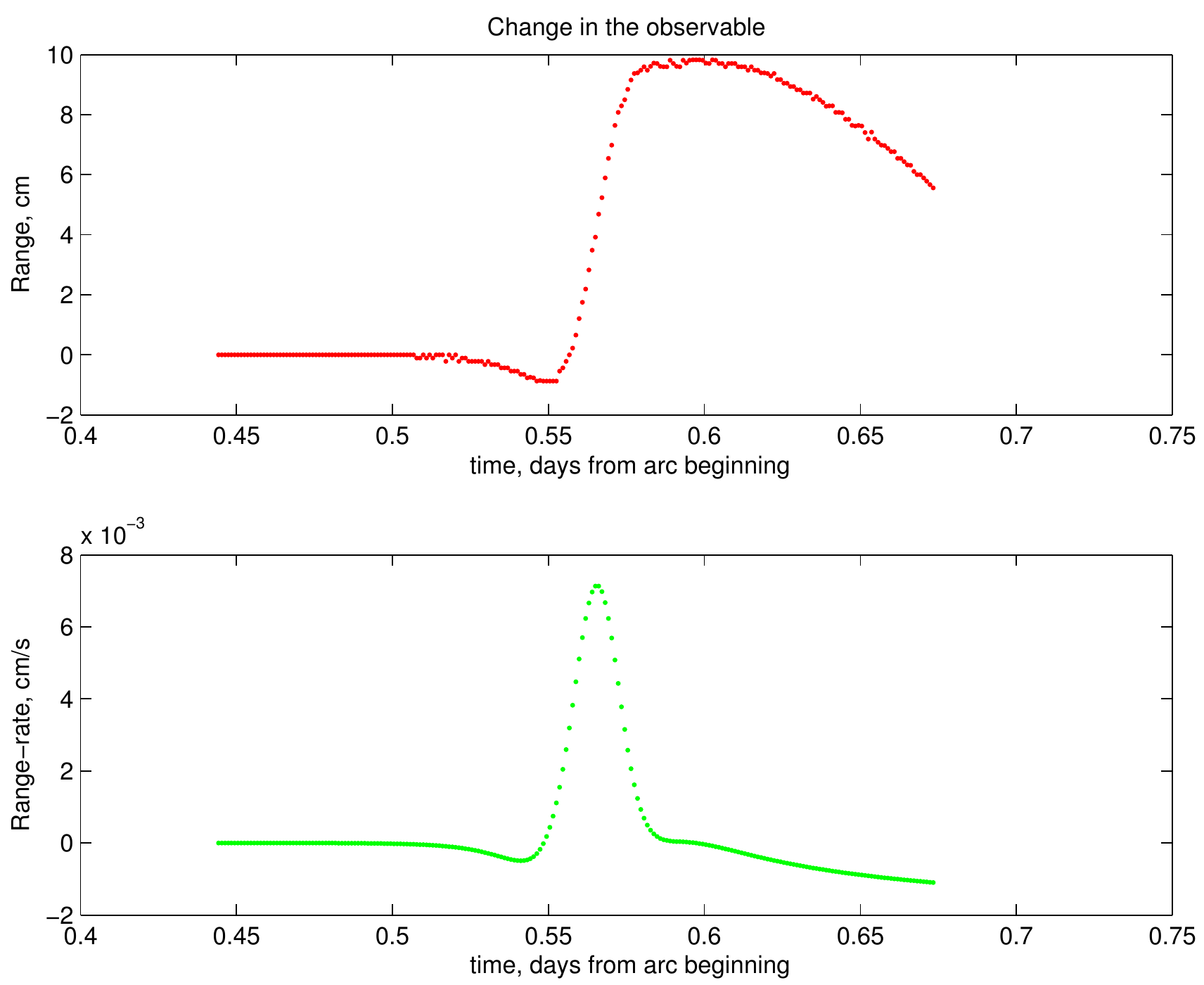}
\caption{Perturbation from the tidal degree 4 term from
all the Galilean satellites and the Sun.}
\label{fig:k4}
\end{figure}

Going on with the exploration of the higher degrees of the solid tide
we arrive to Fig.~\ref{fig:k4}, where the sensitivity of the
observables to the tidal degree $4$ term from all the Galilean
satellites and the Sun is shown (for the simulation $k_4=0.7$ has
been assumed): the $S/N$ is about $20$ and it means that
we could extract information from the data also about $k_4$.

\subsection{Relativistic effects}

The Juno mission has not been designed to improve the knowledge about
GR: a SCE, with the goal of constraining the parameter $\gamma$ has
been performed last summer, but it is fairly certain that the results
cannot improve that obtained with the Cassini mission (\cite{BIT}).
However, the S/C is exposed to significant relativistic effects. In
particular, the high velocity at pericenter ($~60$ km/s), in
combination with Jupiter's fast rotation ($T=~10$ h), induces a
significant acceleration due to the Lense-Thirring (LT) precession. In
the low-velocity, weak field approximation, this acceleration is
proportional to the angular momentum of the central body and to the
velocity of the test particle, and orthogonal to them. A measurement
of the LT precession would therefore provide also the angular momentum
of the planet. As the perturbing field rapidly decreases with the
radial distance, by far the largest acceleration occurs during the
pericenter pass (about $6$ h). This unique opportunity to observe the
LT precession on a planet other than the Earth was first pointed out
in \cite{IO}.

\begin{figure}
\includegraphics[width=7.5cm]{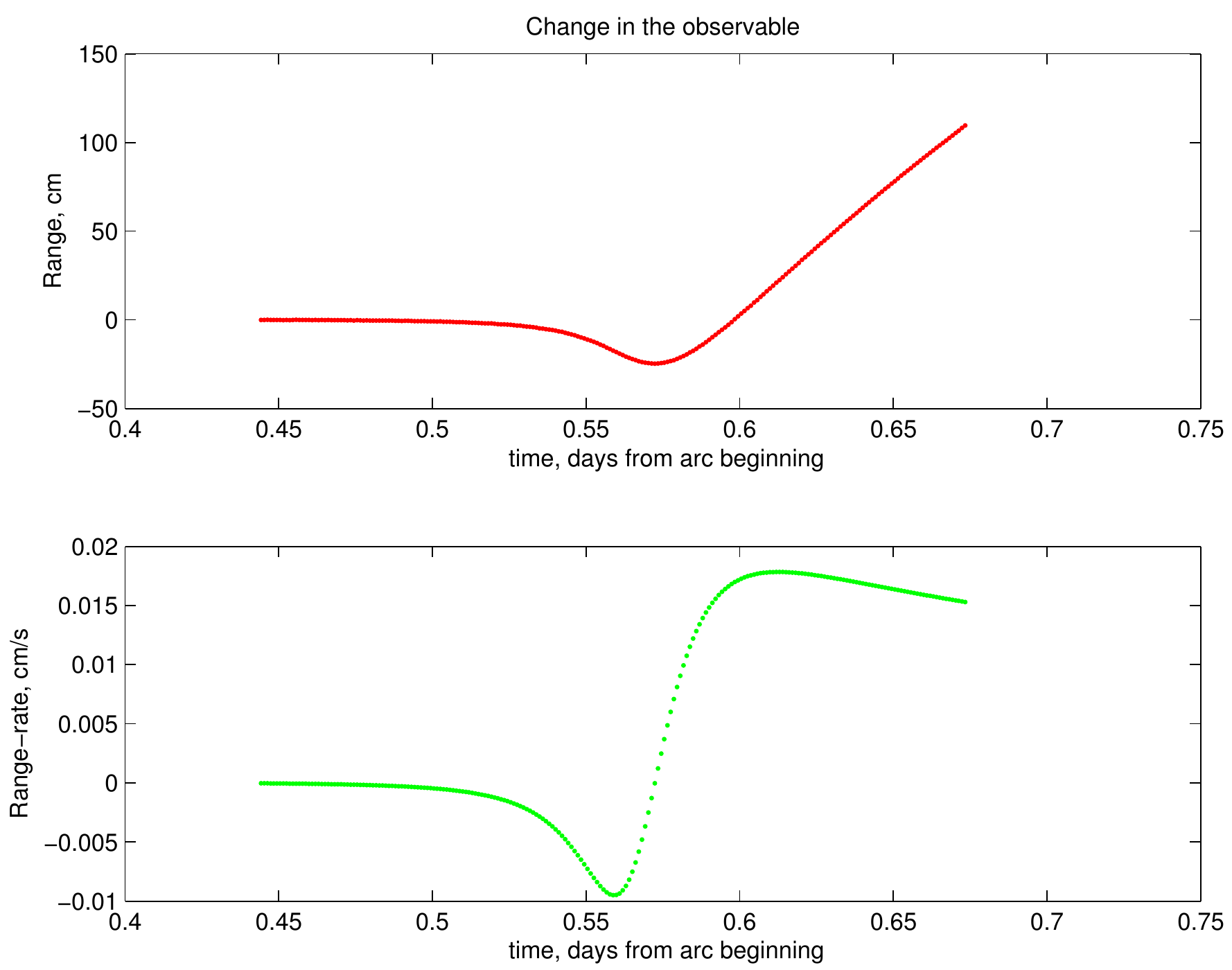}
\caption{Sensitivity of the observables to the Lense-Thirring effect.}
\label{fig:lt}
\end{figure}

In Fig.~\ref{fig:lt} we show the sensitivity of the observables to the
Lense-Thirring relativistic effect (for the simulation, $GJ=2.83\times
10^{38} $ cm$^5$/s$^3$ has been assumed, where $J$ is the angular momentum of Jupiter): 
the $S/N$ is about $100$,
thus we hope to obtain a very significant constraint on the angular
momentum of Jupiter.

\subsection{Gravity field}

\begin{figure}
\includegraphics[width=7.5cm]{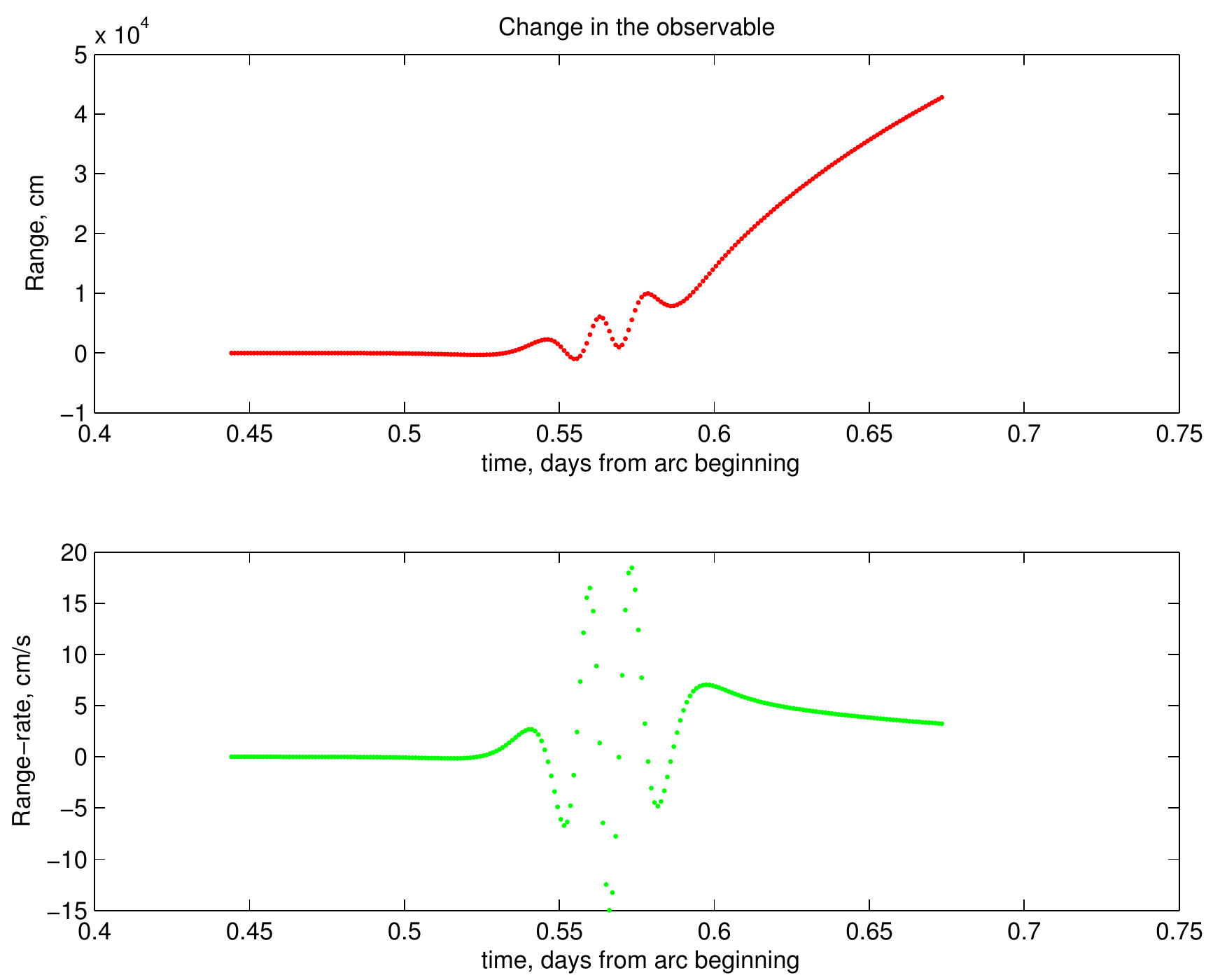}
\caption{Sensitivity of the observables to the
  degree $6$ zonal harmonic $C_{60}$.}
\label{fig:ites46}
\end{figure}

One of the most important goal of the Juno mission is to accurately
map the gravitational field of Jupiter with unprecedented
accuracy. The very low pericenter (about $5000$ km altitude) makes the
orbit especially sensitive to the zonal gravity field. Through our
simulation software we analyzed the impact of zonal harmonics on the
observables: for example, in Fig.~\ref{fig:ites46} we show the
sensitivity to the degree $6$ zonal harmonic $C_{60}$ (for the
simulation, $C_{60}= -3.4\times 10^{-5}$ has been assumed). The
availability of a $S/N\sim 10^5$ does not imply the possibility to
solve for this individual coefficient with a relative accuracy of the
order of $10^{-5}$, although most of the problem shows up at higher
degrees (see \cite{SE}).  The problems at low degrees are caused by
the correlation of the zonal harmonics with the tesseral ones.

\begin{figure}
\includegraphics[width=7.5cm]{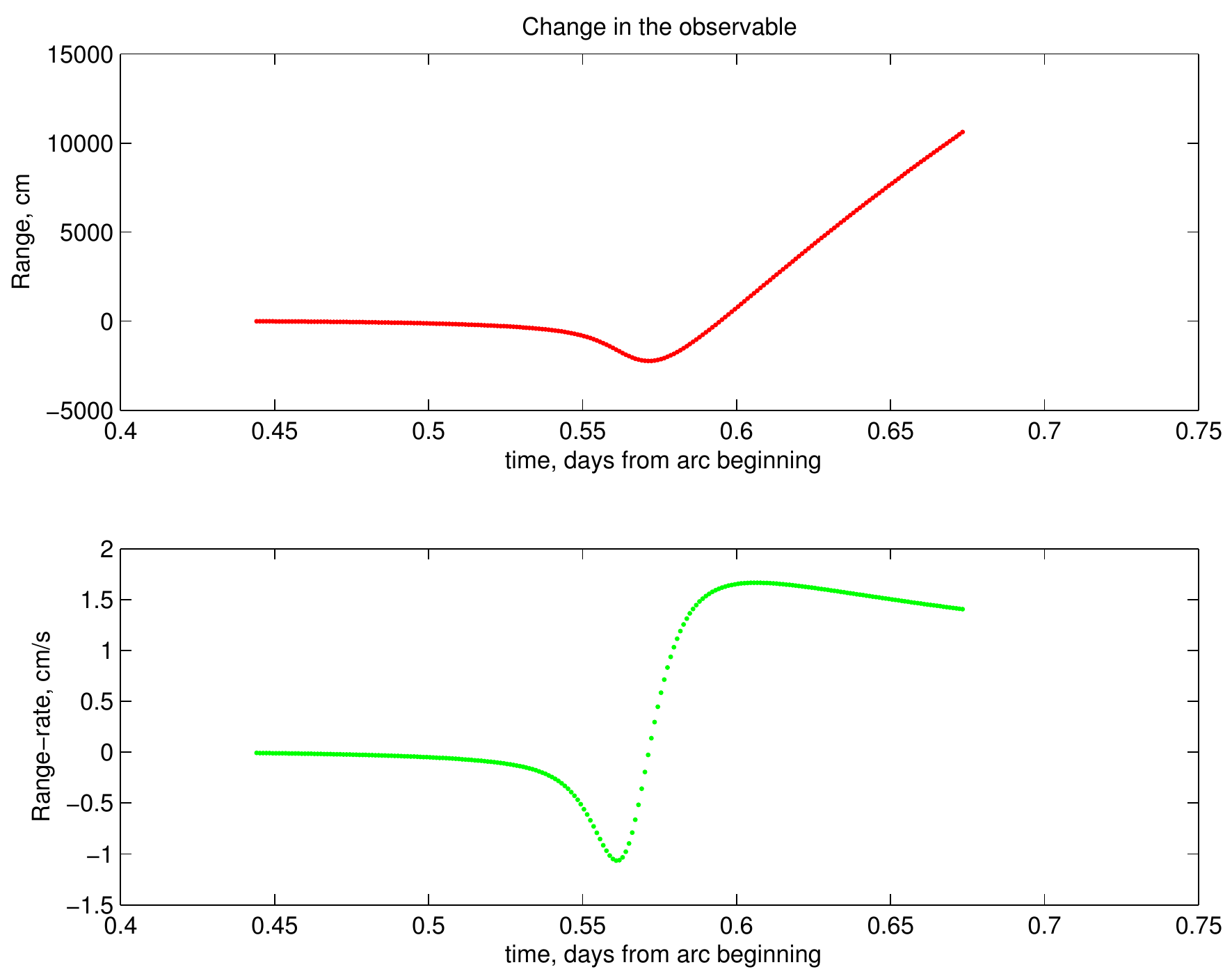}
\caption{Sensitivity of the observables to a change of
  the symmetry axis by $10^{-5}$ radians, that is $~2$ arcsec. }
\label{fig:oblvar1}
\end{figure}

\begin{figure}
\includegraphics[width=7.5cm]{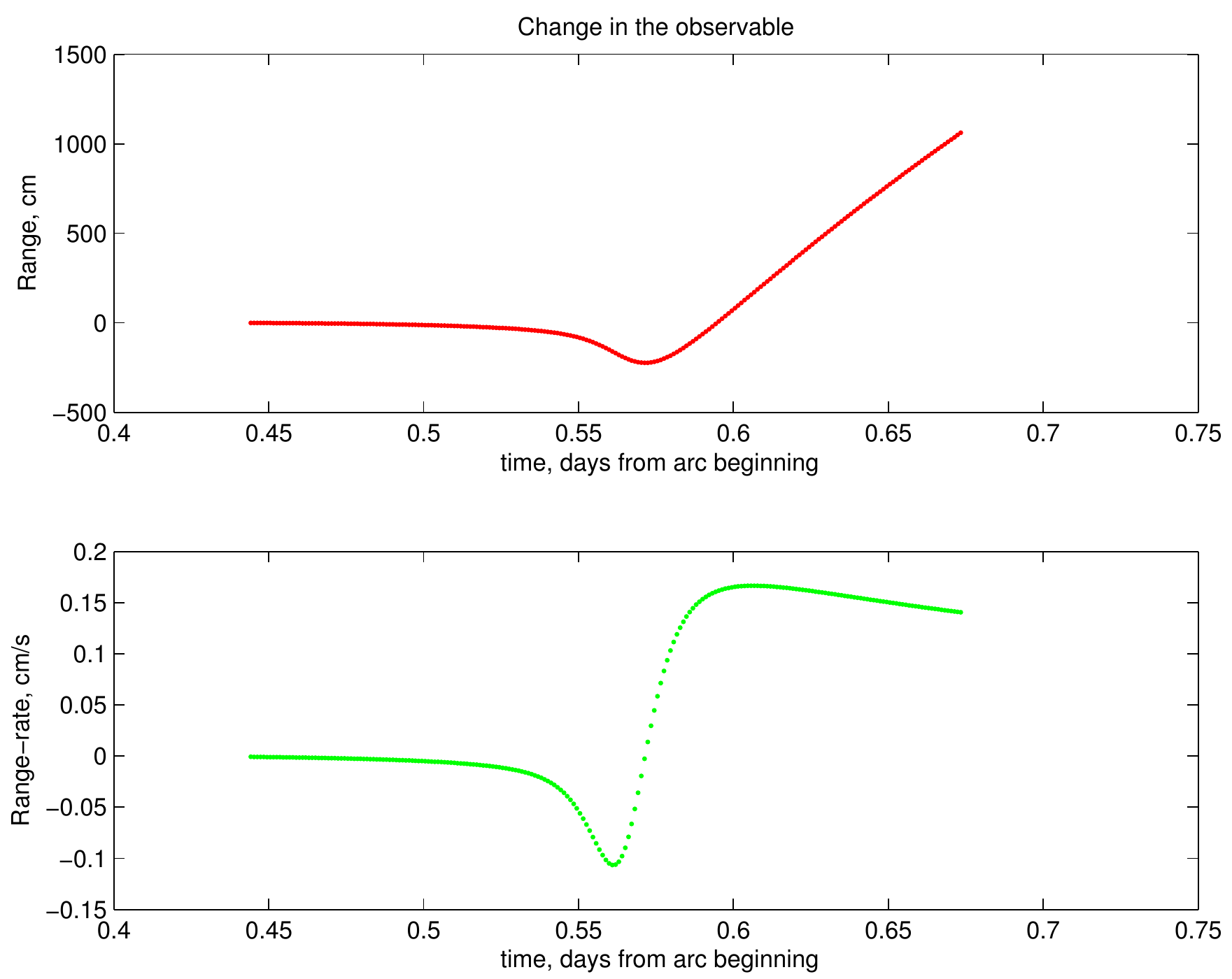}
\caption{Sensitivity of the observables to a change of
  the symmetry axis by $10^{-6}$ radians}
\label{fig:oblvar2}
\end{figure}

\begin{figure}
\includegraphics[width=7.5cm]{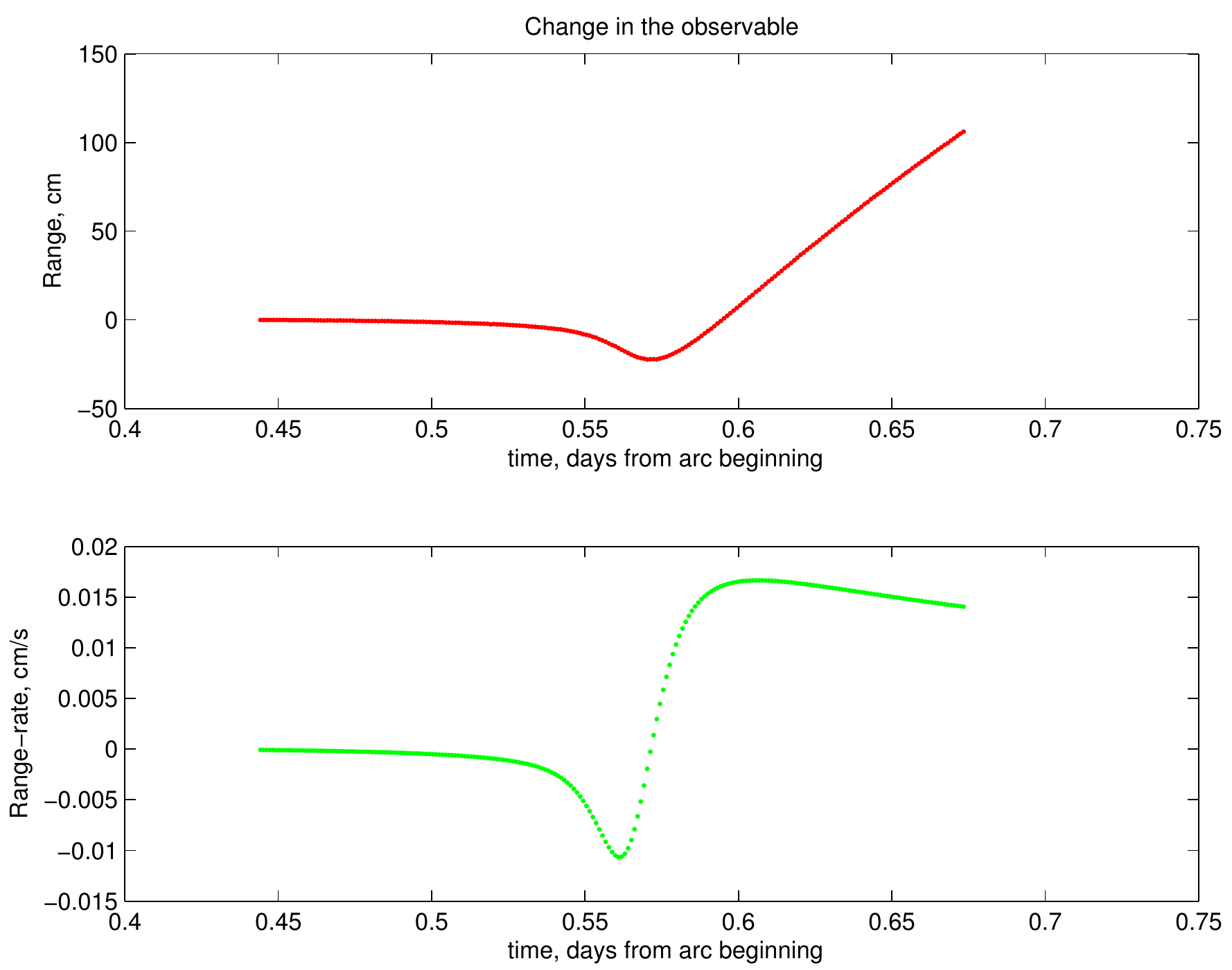}
\caption{Sensitivity of the observables to a change of
  the symmetry axis by $10^{-7}$ radians}
\label{fig:oblvar3}
\end{figure}

Thus, we have to take into account tesseral harmonics: look at
Fig.~\ref{fig:oblvar1}, \ref{fig:oblvar2}, \ref{fig:oblvar3} where we
show the sensitivity of the observables to a change of the symmetry
axis by $10^{-5}$, $10^{-6}$ and $10^{-7}$ radians respectively.  The
large S/N is due to the spill into tesserals, under a small off-axis
rotation, of the zonal harmonics, especially $C_{20}$. Thus, the
accuracy in the symmetry axis is limited by the tesseral harmonics.

\begin{figure}
\includegraphics[width=7.5cm]{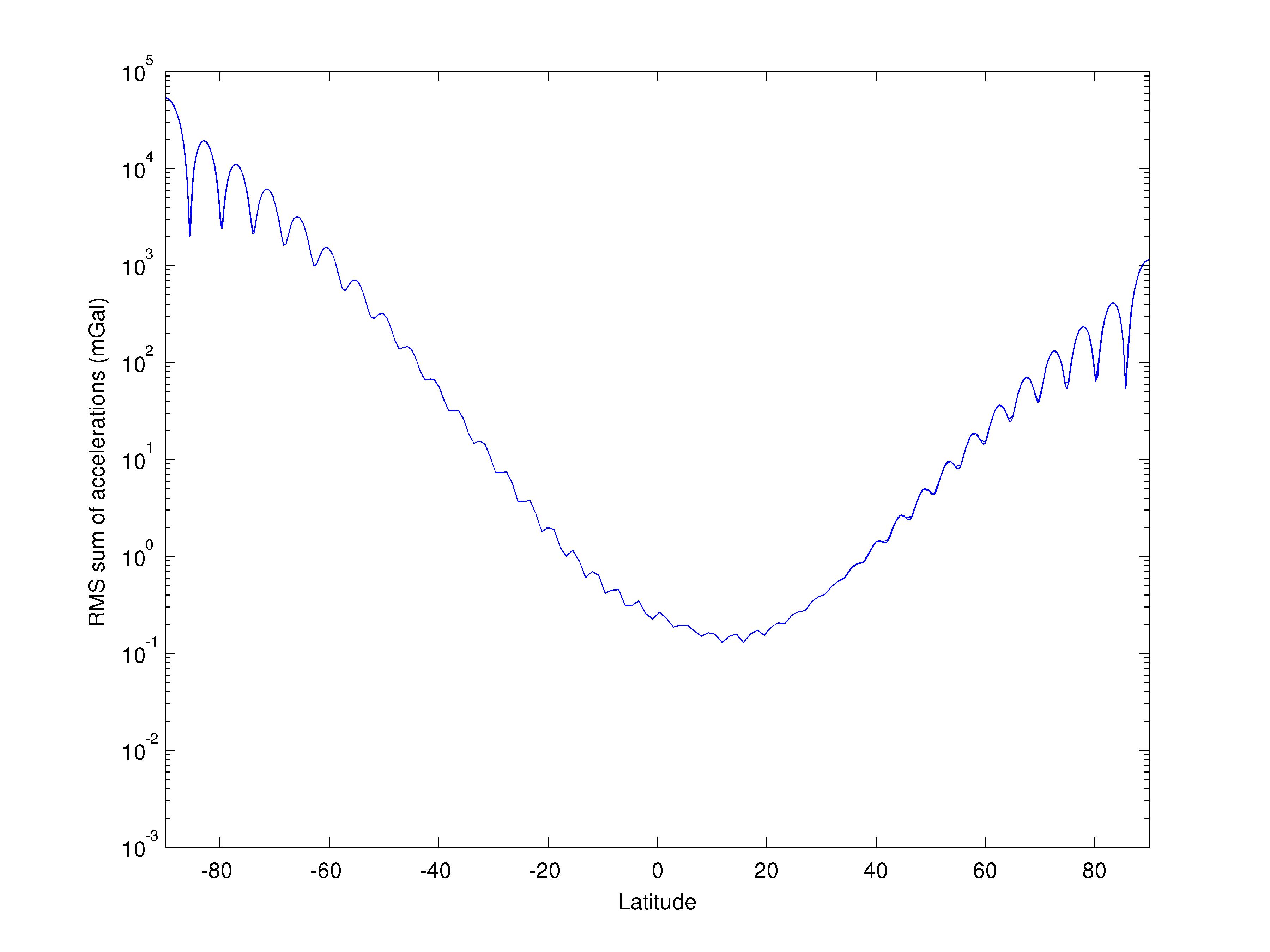}
\caption{Gravity anomalies (semilogarithmic scale).}
\label{fig:gravan}
\end{figure}

From Fig.~\ref{fig:oblvar1}, \ref{fig:oblvar2}, \ref{fig:oblvar3} it
is also clear that tesseral harmonics, even if they are small, are
going to be the main source of irreducible residuals, if they are not
modeled. Thus some of them will have to be solved, possibly with a
priori constraints.  Moreover, the zonal harmonics are strongly
correlated (\cite{KHSF}), thus it will be simply not possible to solve
for each individual zonal harmonic beyond a comparatively low
degree. E.g., for order between $24$ and $30$, the accuracy of the
individual coefficients is going to be of the order of $10^{-4}$,
which is more than the expected signal. Our proposed approach is to
provide a model-independent result, with reliable covariance: this
could take the form of a set of gravity anomalies, limited to the well
observed latitude zone, which is expected to be only $40^\circ$
wide. It is possibile, using a semi-analytic theory, as done in
\cite{SE}, to estimate these gravity anomalies (see
Fig.~\ref{fig:gravan}.  The information about this gravity anomalies,
together with the Love coefficients $k_j$ and with the estimated
angular momentum by the LT effect, should be useful to constrain
models of the interior.

\section{Conclusions and future work}\label{sec:future}

In this work, based on a preliminary but necessary analysis for the Juno RSE, 
we achieved a dual purpose: 
\begin{enumerate}
\item[1)] we described the building of a simulation software, addressing the
dynamical models used;
\item[2)] we showed some preparatory results about the scientific
  goals (gravity field of Jupiter, Love numbers and Lense-Thirring
  relativistic effect) using a sensitivity analysis.
\end{enumerate}

The next step in the Juno RSE software development will be the
implementation of a full least squares fit to the observables to solve
for all the parameters of interest. As described in Sec.~\ref{sec:gs},
there are different ways to do this, and we still have the possibility
of making choices without waste of effort: the main issue to be
discussed with the Juno team is whether we are supposed to use only
the passes at perijove dedicated to the gravity experiment, or also
other tracking sessions, to attempt a model of the intermediate
arcs. Once the corrector software will be developed and tested, we
will proceed to full cycle simulations of the experiment in Jupiter
orbit, with the goal of pushing the performance to its intrinsic
limits. This to some extent depends upon our understanding of the
science goals, and from our capability of handling the mathematical
difficulties of a poorly conditioned orbit determination problem.

\section*{Acknowledgments}\label{sec:ack}

This work has been supported by the Italian Space Agency under the
contract \emph{Radioscienza per BepiColombo e Juno-fasi
  B2/C/D-Attivit\`a scientifiche }. We would like to thank the two anonymous
  reviewers for their suggestions.\\

\end{document}